\documentclass[12pt,preprint]{aastex}
\usepackage{apjfonts,graphicx,graphics}
\usepackage{amsmath}
\usepackage{natbib}
\newcommand{\simless}
     {\ensuremath{\lower
3pt\hbox{$\rlap{\raise5pt\hbox{$\char'074$}}\mathchar"7218$}}}
\newcommand{\simgreat}
     {\ensuremath{\lower
3pt\hbox{$\rlap{\raise5pt\hbox{$\char'076$}}\mathchar"7218$}}}
\newcommand{\simgt}{\lower.5ex\hbox{$\; \buildrel > \over \sim \;$}}
\newcommand{\simlt}{\lower.5ex\hbox{$\; \buildrel < \over \sim \;$}}
\shorttitle{Magnetic Field Strength Maps}
\shortauthors{Koch et al.}
\begin{document}
\title{Magnetic Field Strength Maps for Molecular Clouds:  
A New Method Based on a Polarization - Intensity Gradient Relation}
\author{
Patrick M. Koch\altaffilmark{1},
Ya-Wen Tang\altaffilmark{1,2,3}
\&
Paul T. P. Ho\altaffilmark{1,4}
}
\altaffiltext{1}{Academia Sinica, Institute of Astronomy and
 Astrophysics, Taipei, Taiwan}
\altaffiltext{2}{Universit\'e de Bordeaux, Observatoire Aquitain des Sciences de l'Univers,
2 rue de l'Observatoire, BP 89, F-33271 Floirac Cedex, France}
\altaffiltext{3}{CNRS, UMR 5804, Laboratoire d'Astrophysique de Bordeaux,
2 rue de l'Observatoire, BP 89, F-33271 Floirac Cedex, France}
\altaffiltext{4}{Harvard-Smithsonian Center for Astrophysics, 60
 Garden Street, Cambridge, MA 02138, USA}

\email{pmkoch@asiaa.sinica.edu.tw}
%
%
\begin{abstract}

Dust polarization 
orientations
 in molecular clouds often tend to be close
to tangential to the Stokes $I$ dust continuum emission contours. The magnetic 
field and the emission gradient 
orientations,
therefore, show some correlation.
A method is proposed, which -- in the framework of ideal magneto-hydrodynamics 
(MHD) -- connects the measured angle between magnetic field and emission 
gradient 
orientations
to the 
total
field strength. The approach is based on the 
assumption that a change in emission intensity (gradient) is a measure for 
the resulting direction of motion in the MHD force equation.
In particular, this new method leads to maps of position-dependent magnetic 
field strength 
estimates. When evaluating the field curvature and the gravity direction locally 
on a map, the method can be generalized to arbitrary cloud shapes.
The technique is applied to high-resolution ($\sim0\farcs7$) Submillimeter Array 
polarization data of the collapsing core W51 e2. A tentative $\sim 7.7$~mG field 
strength is found when averaging over the entire core. The analysis further 
reveals some structures
and an azimuthally averaged radial profile $\sim r^{-1/2}$ for the field strength.
Maximum values close to the center are around $19$~mG.
The 
currently available observations lack higher resolution data to probe the 
innermost part of the core where the largest field strength is expected 
from the method. Application regime and limitations of the method 
are discussed. As a further important outcome of this technique, 
the local significance of the magnetic field force compared to the other 
forces can be quantified in a model-independent way, from measured angles only.
Finally, the method can potentially also be expanded and applied to other 
objects (besides molecular clouds) with measurements that reveal the field
morphology, as e.g. Faraday rotation measurements in galaxies. 

\end{abstract}
%
%
\keywords{ISM: clouds --- ISM: magnetic fields, polarization
--- ISM: individual
          (W51 e2) --- Methods: polarization}
%
%
\section{Introduction}     \label{intro}

Magnetic fields are being recognized as one of the key components in star formation 
theories \citep[e.g.,][]{mckee07}. Nevertheless, their exact role in the formation and evolution
of molecular clouds is still a matter of debate in the literature. On the 
observational side, one of the difficulties is to accurately measure the 
magnetic field strength.
The Zeeman effect still provides the only known method of directly measuring
magnetic field strengths along the line of sight in a molecular cloud
\citep[e.g.,][]{troland08, crutcher09}.
A statistical method, aiming at separating a three-dimensional uniform field
and a non-uniform field component is given in \citet{jokipii69} and was later
extended and applied to optical polarization data in \citet{myers91}.
The Goldreich-Kylafis effect, the linear polarization in molecular line spectral
emission \citep{gk81,gk82}, is a mechanism to trace the magnetic
field morphology. Up to now, the field structures of only a few sources have
been detected with this effect \citep[e.g.,][]{cortes08,beuther10}.
Polarization of dust thermal emission at infrared and submillimeter
wavelengths provides another method to study magnetic field properties
\citep[e.g.,][]{hildebrand00}.
The dust grains are thought to be aligned with their shorter axis
parallel to the magnetic field lines. Therefore, the emitted light 
appears to be polarized perpendicular to the field lines
\citep{cudlip82,hildebrand84,hildebrand88, lazarian00}.
Radiative torques are likely to be responsible for the dust alignment
\citep{draine96, draine97, lazarian00, cho05, lazarian07}.

Dust polarization observations have been obtained both for 
cores together with their surrounding diffuser material
\citep[e.g.,][]{schleuning98, lai01}
and 
for higher-resolution
collapsing cores \citep[e.g.,][]{girart06, tang09}.
Complementary to Zeeman splitting, dust polarization measurements are
probing the plane of sky projected magnetic field structure. They, however, 
do not directly yield a magnetic field strength, but only some field
morphology. Additional 
modeling is needed in order to derive the actual field strength 
perpendicular to the line of sight. 
A three-dimensional model of an hourglass magnetic field configuration
is fitted to dust observations in, e.g. \citet{kirby09} to yield a mean field
strength.
With pinched field configurations detected, comparing gravitating 
force and field tension gives an upper limit for the field strength
\citep[e.g.,][]{schleuning98, tang09, rao09}.
Another commonly used technique is the 
Chandrasekhar-Fermi (CF) method \citep{chandra53}, where a field 
strength is derived from the field dispersion (around a mean field direction)
combined with a turbulent velocity information. Various modifications and corrections
to this method are discussed in the literature 
\citep[e.g.,][]{zweibel90, houde04b, falceta08}
and tested through magneto-hydrodynamics (MHD) simulations
\citep{ostriker01,padoan01,heitsch01,kudoh03}.
The work by \citet{hildebrand09} further refines this method -- 
noting that a globally derived dispersion might not reflect the true
contribution from magneto-hydrodynamic waves and/or turbulence --
by introducing the concept of a dispersion function (structure function)
around local mean magnetic field 
orientations.

However, the CF method and all its variations are statistical in 
nature, relying on (large) ensembles of polarization segments. 
Thus, they typically provide one {\it ensemble-averaged} value for the magnetic
field strength for an entire map. Consequently, zones of weaker and 
stronger field strengths can not be separated. It is, therefore, non-trivial
to evaluate the relative, generally position-dependent role of 
the magnetic field (compared to, e.g., gravity, turbulence).
The goal of this paper is to introduce a new method which leads to a 
{\it local} magnetic field strength and, therefore, can map the 
position-dependent field strength over an entire molecular cloud

The paper is organized as follows.
Section \ref{source} summarizes the Submillimeter Array (SMA) polarization
observation of the source W51 e2 which serves as an illustration case.
Section \ref{method} introduces our new method, starting with an observed
correlation, the dust - MHD connection and a qualitative analysis of 
the predicted features of the method.
A possible extension of the method to an arbitrary cloud shape is 
discussed in Section \ref{general}. The results are presented
in Section \ref{results}. We follow up with a discussion in 
Section \ref{discussion} which addresses application regime and limitations
of the method, and which also proposes a measure to evaluate the 
dynamical role of the magnetic field in different zones of a molecular
cloud. Summary and conclusion are given in Section \ref{summary}.

\section{An Illustration Case: W51 e2} \label{source}

The W51 e2 map (Figure \ref{w51_e2}) analyzed in the following 
sections was obtained with the SMA interferometer \citep{tang09}. 
The observation was carried out 
at a wavelength of 0.87~mm where the polarization of dust thermal emission 
is traced. In the extended array configuration, the synthesized beam 
resolution was about $0\farcs7$ with a primary beam (field of view)
of about $30\arcsec$ at 345~GHz.
A detailed description of the data analysis with a discussion and 
interpretation of the 
obtained images is given in \citet{tang09}.
The re-constructed features depend on the range of $uv$-sampling
and weighting.
Nevertheless, these data currently provide the highest angular resolution
($\theta$) information
on the morphology of the magnetic field in the plane of sky obtained
with the emitted polarized light in the W51 star formation site. 
This study is part of the program on the SMA\footnote{
The Submillimeter Array is a joint project between the Smithsonian
Astrophysical Observatory and the Academia Sinica Institute of Astronomy
and Astrophysics, and is funded by the Smithsonian Institution and the
Academia Sinica.
}
\citep{ho04} to study the structure of the magnetic field at
high spatial resolutions.

W51 e2 is one of the strongest mm/submm continuum sources in the W51
region.
Located at a distance of 7 kpc \citep{genzel81},
1$\arcsec$ is equivalent to $\sim$ 0.03 pc.
\citet{chrys02} measured the polarization at 850$\mu$m with SCUBA on
JCMT across the molecular cloud at a scale of 5 pc with
a  binned resolution $\theta\approx 9\farcs$3.
At this scale, the polarization 
shows a morphology that changes from the dense cores to the 
surrounding diffuser gas.
\citet{lai01} reported a higher angular resolution (3$\arcsec$)
polarization map
at 1.3 mm obtained with BIMA, which resolved out large scale structures.
In contrast to the larger scale polarization map,
the polarization appears to be uniform across the envelope at a scale of
0.5 pc,
enclosing the sources e2 and e8. In the highest angular resolution map
obtained
with the SMA at 870 $\mu$m with $\theta \approx 0\farcs$7,
the polarization patterns appear to be pinched in e2 and also possibly in
e8 \citep{tang09}. In particular, the structure detected in the collapsing
core e2 
reveals hourglass-like features. 
A statistical analysis based on a polarization structure function (of second
order) shows a turbulent to mean magnetic field ratio which decreases
from the larger (BIMA) to the smaller (SMA) scales from $\sim 1.2$ 
to $\sim 0.7$ \citep{koch10},
possibly demonstrating that the role of magnetic field and turbulence
evolves with scale.
The apparent axis of the hourglass along the north-west south-east direction
coincides with the axes of the bipolar outflows in CO(2-1) \citep{keto08},
CO(3-2) and HCN(4-3) \citep{shi10}. 
Collapse and/or accretion flow signatures, possibly orthogonal to that, have been reported for 
various molecules in \citet{rudolph90, ho96, zhang97, young98, zhang98, solins04,keto08}.
This might correspond roughly to the depolarization regions along the 
north-east south-west direction in Figure \ref{w51_e2}.

The highest resolution SMA map of W51 e2 
serves as a testbed for the method developed in the following section.
Figure \ref{w51_e2} reproduces the dust continuum Stokes $I$ map
with the magnetic field segments (red) from \citet{tang09}. Detected 
polarization segments are rotated by 90$^{\circ}$ in order to derive 
the field directions.
Typical measurement uncertainties of individual position angles are in the 
range of $5^{\circ}$ to $10^{\circ}$. Additionally shown in the map 
are features relevant for the method in the following section.

\section{Method} \label{method}

In this section we introduce a new method which allows us to measure the (projected)
magnetic field strength in a molecular cloud as a function of the position 
in a map. Starting from an 
observed correlation in Section \ref{motivation}, the details of the 
method and its link to observed dust continuum Stokes $I$ and polarization maps
are given in Section \ref{method_detail}.
We follow up with a qualitative analysis of the features of the method in 
Section \ref{analysis}.

\subsection{Motivation: Magnetic Field -- Intensity Gradient Correlation} 
            \label{motivation}

The Stokes $I$ dust continuum contours and the magnetic field segments (red)
from the dust polarization emission \citep{tang09} are shown in Figure 
\ref{w51_e2} for W51 e2. Interestingly, field segments tend to be perpendicular
to Stokes $I$ emission contours, or equivalently, polarization is preferentially 
tangential to the emission contour lines. Instead of further working with intensity 
contours, we alternatively choose to investigate the intensity gradients 
(blue thick segments in Figure \ref{w51_e2}).
The top panel in
 Figure \ref{correlation_b_i_grad} displays the magnetic field position angles ($P.A.$s)
versus the intensity gradient $P.A.$s. A close correlation becomes manifest. Both $P.A.$s
are defined counter-clockwise starting from north in a range of 0 to 180$^{\circ}$. 
The few data points (black empty squares) which originally 
do not seem to follow the trend, belong to pairs where both segments are 
close to $P.A.=0$ (vertical 
orientation),
but with one of the $P.A.$s 
slightly rotated to 
the left and 
the other one 
slightly rotated to 
the right hand side of the vertical. 
Despite being closely correlated in orientation (small absolute differences), these
pairs would mimic a bad correlation simply due to the definition of the $P.A.$s. 
For these cases, the original range of the magnetic field $P.A.$ 
(0 to 180$^{\circ}$) is extended by adding their relative difference in $P.A.$s
to their intensity gradient $P.A.$s. E.g., a pair with an intensity gradient
and magnetic field $P.A.$ of 175$^{\circ}$ and 5$^{\circ}$ has a relative
difference of 10$^{\circ}$, and is therefore displayed as (175$^{\circ}$,185$^{\circ}$).
This re-definition does not alter the correlation in any way, but is merely
a consequence of the ambiguity in the conventional definition of the $180^{\circ}$
$P.A.$ range.

We note that the correlation is not limited to those pairs directly 
in the core collapsing region, 
but seemingly still holds for the most northern segments where the
emission contours start to deviate from circular shapes.
The correlation -- in the definition of Pearson's linear correlation 
coefficient -- is 0.95. It is important to remark that, despite this tight
correlation, the distribution of the differences (between magnetic field $P.A.$s
and intensity gradient $P.A.$s) is non-Gaussian with an
absolute
mean of $\sim 20^{\circ}$. 
A KS-test rejects the null 
hypothesis (differences following a Gaussian distribution) with a $p-$value
of $\sim 10^{-7}$. 
As a matter of fact, the relative differences show a bimodal distribution 
(bottom panel in Figure \ref{correlation_b_i_grad}), which results from a 
systematic clockwise and counter-clockwise rotation of the magnetic 
field with respect to the intensity gradient. This further points toward
a systematic origin of this correlation.
All this leads to the conclusion that magnetic field and 
intensity gradient can not simply be radially aligned with the apparent 
misalignment just being the result of some measurement uncertainty.
It is the focus of an on-going study where this correlation is further analyzed on a 
larger data sample in the context of the evolution of molecular clouds
(Koch et al. 2012; in preparation).

In the following section, this observed correlation and the measured non-Gaussian
differences in position angles of the magnetic field and the intensity
gradient are further used to investigate the field strength.

\subsection{MHD and Dust Polarization}   \label{method_detail}

\subsubsection{Local Field Strength}

Molecular clouds are the sites of interactions of various forces.
Gravity, magnetic fields, expanding shells such as from expanding HII regions
or supernovae, accretion and outflows all
add to the dynamics, and their relative importance can vary with the 
evolutionary stage of the star formation site. Generally, magneto-hydrodynamics
(MHD) equations - possibly for various molecules and dust with individual coupling 
terms - are needed to provide an accurate and complete description of 
the system. In the following, aiming at making use of dust polarization 
measurements, a single MHD equation for the dust component is adopted\footnote
{Simplifying and describing the molecular cloud dynamics with a single 
equation relies on the assumption of (local) collisional equilibrium 
between dust and other particles. Since the dust particles both couple
to the magnetic field and also follow gravity and any other pressure gradient, 
other species will in a first approximation follow the same dynamics.
}. 
Under the assumptions of negligible viscosity and infinite conductivity
(ideal MHD case) we focus on the force equation:
\begin{equation}
\rho\left(\frac{\partial}{\partial t}+\mathbf{v}\cdot\mathbf{\nabla}\right)\mathbf{v} =
-\mathbf{\nabla}\left( P + \frac{B^2}{8\pi}\right)-\rho\nabla\phi+\frac{1}{4\pi}
\left(\mathbf{B}\cdot\mathbf{\nabla}\right)\mathbf{B},            \label{mhd_momentum}
\end{equation}
where $\rho$ and $\mathbf{v}$ are the dust density and velocity, respectively.
$\mathbf{B}$ is the magnetic field. $P$ is the hydrostatic dust pressure. 
$\phi$ is the gravitational potential resulting from the 
total mass contained inside the star forming region. $\mathbf{\nabla}$ denotes
the gradient. As usual, the left hand side in Equation (\ref{mhd_momentum}) 
describes the resulting action based on the force terms on the right hand side.
These include the 
gradients of the
hydrostatic pressure terms of the gas, the magnetic field
and the gravitational potential together with the magnetic field tension
term (last term on the right hand side).

In the following, an interpretation of Equation (\ref{mhd_momentum}) is 
derived which can be matched to a polarization observation as in Figure \ref{w51_e2}.
We start by noting that the magnetic field tension term, 
$\sim (\mathbf{B}\cdot\mathbf{\nabla})\mathbf{B}$, can be rewritten as:
\begin{equation}
\left(\mathbf{B}\cdot\mathbf{\nabla}\right)\mathbf{B}
=B\frac{\partial B}{\partial s_B}\mathbf{e}_{s_B}+B^2\frac{\partial\mathbf{e}_{s_B}}{\partial s_B}
=B\frac{\partial B}{\partial s_B}\mathbf{e}_{s_B}+B^2\frac{1}{R}\mathbf{n},  \label{B_curvature}
\end{equation}
where a magnetic field line, $\mathbf{B}=B\mathbf{e}_{s_B}$, is directed
along the unity vector $\mathbf{e}_{s_B}$
with the generalized coordinate $s_B$ along the field force line.
A change in direction, 
$\partial\mathbf{e}_{s_B} / \partial s_B$, measures the magnetic field curvature $1/R$ and 
is directed normal to $\mathbf{e}_{s_B}$ along the unity vector $\mathbf{n}$.
In an analogous way, $(\mathbf{v}\cdot\mathbf{\nabla})\mathbf{v}$ can be 
reformulated with the velocity vector $\mathbf{v}=v\mathbf{e}_{s_v}$
directed along the unity vector $\mathbf{e}_{s_v}$. 

Generally, the gradient of the isotropic field pressure, $\sim\nabla\left(\frac{B^2}{8\pi}\right)$,
in Equation (\ref{mhd_momentum}) is not necessarily directed along the field
direction $\mathbf{e}_{s_B}$, but can also have a component orthogonal to it.
Therefore, when combining the Equations (\ref{B_curvature}) and (\ref{mhd_momentum}),
a term $\sim B\frac{\partial B}{\partial s_{B_{\bot}}}\mathbf{e}_{s_{B_{\bot}}}$ remains
which needs to be compared to the curvature term $\sim B^2\frac{1}{R}\mathbf{n}$. 
Both $\mathbf{n}$ and $\mathbf{e}_{s_{B_{\bot}}}$ are orthogonal to $\mathbf{e}_{s_B}$, 
but they are not necessarily collinear. Assuming $\partial s_{B_{\bot}}$ and $R$ to be given
approximately by the resolution of an observation, the significance of this 
additional term can be estimated as 
$| B\frac{\partial B}{\partial s_{B_{\bot}}}|/\left(B^2\frac{1}{R}\right)\approx \frac{\Delta B_{\bot}}{B}\ll 1$.
Thus, the term can be omitted if the local change in the normal field component, $\Delta B_{\bot}$,
is small compared to the total field strength. This holds for any smooth and 
slowly varying function of field strength.
Equation (\ref{mhd_momentum}) can then be transformed into:
\begin{equation}
\rho v \frac{\partial v}{\partial s_v}\mathbf{e}_{s_v}
+\rho v^2 \frac{\partial\mathbf{e}_{s_v}}{\partial s_v}
= -\frac{\partial P}{\partial s_P}\mathbf{e}_{s_P} 
- \rho\frac{\partial\phi}{\partial s_{\phi}}\mathbf{e}_{s_{\phi}}
+\frac{1}{4\pi}B^2\frac{1}{R}\mathbf{n},                         \label{mhd_2}
\end{equation}
where the resulting directions of the gradients of pressure and gravitational
potential are given with $\mathbf{e}_{s_P}$ and $\mathbf{e}_{s_{\phi}}$, respectively.
In the above formulation, generalized coordinates $s_v, s_P, s_{\phi}, s_B$ 
along the directions of the unity vectors $\mathbf{e}_{s_v}, \mathbf{e}_{s_P},
\mathbf{e}_{s_{\phi}}, \mathbf{e}_{s_B}$ are used. 
Generally, these vectors are not 
collinear, but have different three-dimensional directions with the condition
that the vector sum of the right hand side adds up to the resulting direction
on the left hand side. Each term in Equation (\ref{mhd_2}) is intentionally
written with its corresponding unity vector, defining its specific direction.
This will later be used when interpreting Equation (\ref{mhd_2})
together with an observed map. Equivalently, all unity vectors can be 
expressed in any regular coordinate system (e.g. spherical coordinates $r$, 
$\theta$, $\phi$) should this be of advantage. 
The unity vectors $\mathbf{e}_{s_v}, \mathbf{e}_{s_P}, \mathbf{e}_{s_{\phi}}, 
\mathbf{e}_{s_B}$
can be interpreted as volume averaged directions, where the size of a 
volume element will depend on the resolution of an observation.
In the above derivation we have neglected the partial time derivative, 
$\partial /\partial t$, assuming stationarity.

Equation (\ref{mhd_2}) describes the basic interaction of forces and the 
resulting motion. Linking some of these terms (in direction and strength)
with observations will then allow us to solve for others. Our goal is to 
isolate the magnetic field strength $B$ in the term 
$\frac{1}{4\pi}B^2\frac{1}{R}\mathbf{n}$. Therefore, the remaining terms
in Equation (\ref{mhd_2}) need to be identified and characterized. Dust in 
star formation sites is reacting to all the force terms in Equation (\ref{mhd_2}).
Maps of observed dust emission are reflecting the overall result of 
gravity, pressure and magnetic field. Consequently, they represent a 
measure of the resulting motion, the left hand side in Equation (\ref{mhd_2}).
We make the fundamental assumption that a change in emission intensity
is the result of the 
transport of matter driven by the combination of the 
above mentioned forces. Adopting 
this, it then follows that the gradient in emission intensity defines
the resulting direction of motion on the left hand side in Equation (\ref{mhd_2}).
For the directions of the gradients of the pressure and the gravitational
potential we assume here, for simplicity, a spherically symmetrical 
molecular cloud, where the center can be identified from the peak 
emission. This assumption is relaxed in Section \ref{general}, where the method
is generalized to an arbitrary cloud shape.

It is important to remark that the derivation so far is generally valid
for 3 dimensions. Projection effects in both the integrated Stokes $I$
and the polarized emission inevitably are present in an observation.
In the following we first proceed by identifying Equation (\ref{mhd_2})
with an observed map in 2 dimensions. Projection effects are later 
addressed in Section \ref{projection}.
Figure \ref{schematic_b} illustrates the further steps. 
The derivation of $B$ is shown
for a measured polarization (red) and intensity gradient direction (blue)
in the first quadrant (with respect to the gravity center which here
is supposed to coincide with the emission peak). For each measured pair
of polarization and intensity gradient, its distance and direction from 
the gravity center defines the gravitational pull,
$\rho\frac{\partial\phi}{\partial s_{\phi}}\mathbf{e}_{s_{\phi}}$.
Generally, gravity, magnetic field and the intensity gradient directions
are different. The schematic in Figure \ref{schematic_b} illustrates how the vector sum
of the various terms in Equation (\ref{mhd_2}) is constructed. 
Solving for the magnetic field term then relies on measurable angles in 
the 
orientations
between polarization and the intensity gradient ($\alpha$)
and the difference between the gravity and the intensity gradient 
directions ($\psi$). Applying a sinus theorem to the closed triangle, 
$\frac{\nabla P+\rho\nabla\phi}{\sin \alpha}=\frac{\frac{1}{4\pi} B^2 \frac{1}{R}}{\sin\psi}$,
leads to the expression for the magnetic field strength:
\begin{equation}
B=\sqrt{\frac{\sin\psi}{\sin\alpha}\left(\nabla P+\rho\nabla\phi\right)4\pi R},  \label{B}
\end{equation}
where, in the most general case, all variables are functions of positions in a map.

We note that -- although illustrated for a simple, close to spherically symmetric 
case in Figure \ref{schematic_b} with $\nabla P$ and $\rho\nabla\phi$ aligned
and pointing toward the same center -- Equation (\ref{B}) is generally 
valid for any directions of $\nabla P$ and $\rho\nabla\phi$. Thus, the above
derivation is not restricted to a 
spherical
collapsing core, but can be applied to 
any configuration where the various local force directions can be 
identified (see also Sections \ref{general} and \ref{application}).

\subsubsection{Local Field Significance}   \label{local_field_importance}

As an important further outcome of the method, the angle factor 
$\frac{\sin\psi}{\sin\alpha}$ in Equation (\ref{B}) has a direct physical 
meaning. With the magnetic field 
tension 
force term $F_B=\frac{B^2}{4\pi R}$ and the 
gravitational and pressure forces $|F_G+F_P|=|\rho\nabla\phi + \nabla P|$, 
Equation (\ref{B}) can be reformulated as:
\begin{equation}
\left(\frac{\sin\psi}{\sin\alpha}\right)_{local}=\left(\frac{F_B}{|F_G+F_P|}\right)_{local}
                                                                    \equiv \Sigma_B,
\end{equation}
where we have introduced $\Sigma_B$ to define the field significance. 
This directly quantifies the local impact of the magnetic field force 
compared to the other forces involved (gravity and pressure gradient).
Remarkably, the ratio is free of any model assumptions and simply relies on 
the two measured angles. 
It is purely based on the geometrical imprint by the various forces
left in the field and intensity gradient morphologies.
The angle factor, thus, also provides a model-independent 
qualitative criterion whether the magnetic field can prevent an area in a 
molecular cloud from gravitational collapse $(\Sigma_B>1)$ 
or not $(\Sigma_B<1)$.
The field significance $\Sigma_B$ and its implications for the mass-to-flux
ratio and the star formation efficiency are investigated in \citet{koch11}.

\subsection{Qualitative Analysis}  \label{analysis}

We briefly illustrate the expected features of our method when applied
to a representative field and intensity configuration in a molecular cloud
(Figure \ref{qualitative}). A summary is given in Table \ref{table_analysis}.
Field lines like (I) and (II) are expected in 
a core collapse when gravity is dragging inward the magnetic field which
is flux-frozen to the dust particles. In the case of (I), the field line
is close to the pole, little bent, and the gas is mostly moving along 
the field. Intensity gradients are mostly radial, pointing toward the 
center of gravity. Therefore, in the outer region (a), $R\approx$ const, 
$\delta$ small ($\sin \alpha\approx 1$), $\psi$ small ($\sin\psi\approx 
\psi\approx$ const) and thus $\Sigma_B\approx$ const. 
The field strength $B$ will scale as $\sim r^{-\kappa}$, where $\kappa >0$
describes a power-law profile resulting from density and gravitational
potential gradient (Section \ref{result_field_strength}). 
Small (random) changes in angles are possible due to local turbulence
and can lead to some scatter. In the inner part (b), $R$ might slightly decrease
with the other parameters remaining roughly constant. This will lead to 
similar scalings for $\Sigma_B$ and $B$.
A field line of type (II) is expected around the accretion direction. 
The outer field (a) is still little bent
($R\approx$ const), $\delta\approx$ const and small, $\psi$ small, and
therefore $\Sigma_B\approx$ const and $B \sim r^{-\kappa}$ as in (Ia).
In (b) the field lines start to bend more
significantly ($R$ decreases) and $\delta$ increases ($\sin\alpha$ decreases).
$\psi$ is small ($\approx$ const) or possibly increasing if a a non-spherical
central core is being formed. 
Thus, $\Sigma_B$ can increase as $\sim \frac{1}{\sin\alpha}$. The field 
strength is additionally modulated by $\sqrt{\frac{R}{\sin\alpha}}$ which can lead to 
a steeper or shallower profile than $r^{-\kappa}$ depending on the exact values of $R$ 
and $\alpha$. 
In the most inner section (c), the field 
radius is minimized, field and intensity gradient directions are close to 
orthogonal to each other ($\delta\sim \frac{\pi}{2}$ modulo some turbulent
dispersion, $\alpha\sim 0$), and $\psi$ is small. 
The force ratio $\Sigma_B$ can then possibly further increase locally. 
The field strength is 
thus mostly controlled by $\sqrt{\frac{1}{\sin\alpha}}$ and the power-law exponent $\kappa$,
which will lead to highest values here. Overall, for (I) and (II) we thus 
expect the field strength to generally increase with smaller radius toward the center.

Additional features are seen in Figure \ref{w51_e2}, which are possibly 
not directly linked to the above cases. Region (III) -- corresponding to 
the NE patch around the offset (-0.3,1.8) -- shows a group of parallel
B field segments with parallel intensity gradients. In this outer region
the collapse is likely initiated, but does not yet reveal clear signatures.
$\psi$ is changing, $\delta\approx$ const ($\sin\alpha\approx$ const), $R\approx$
const, and $\Sigma_B$ can therefore increase or decrease. 
The field strength will generally be small here because of the low density.
The relative field significance can still be important here, as gravity has not 
yet fully taken control to shape and align intensity gradient and field 
direction. Area (IV) corresponds to the most northern part in Figure
\ref{w51_e2}. This region is possibly disconnected from the main core, 
likely belonging to another smaller core being formed here. The $\Sigma_B$ 
values are dominated by large changes in $\psi$ when linked to the 
gravitational center of the main core. Due to the lack of a clear 
identification of a local gravity center (in which case $\psi$ would be 
much smaller), $\Sigma_B$ is probably overestimated for the two most 
northern segments.  
Possible solutions to this issue are addressed in Section \ref{local_gravity}.
The field strength is again small here due to the low density.

\section{Generalizing to Arbitrary Cloud Shapes}  \label{general}

This section focuses on the 2 remaining parameters in Equation (\ref{B}), 
which are not yet explicitly locally evaluated: the magnetic field 
radius $R$ (curvature $C\equiv 1/R$) and the gravitational potential $\phi$.
In particular, the initial assumption of a spherically symmetrical 
molecular cloud from Section \ref{method} is given up. It has to be
stressed that the method in Section \ref{method} derived from the 
Equation (\ref{mhd_momentum}) remains unchanged, and is valid for both 
average or locally precise values in $R$ and $\phi$.

\subsection{Local Curvature}        \label{local_curvature}

From the derivation in Section \ref{method}, both angles $\alpha$ and 
$\psi$, the gravitational potential and the pressure $P$ are generally functions
of positions $(x,y)$ in a projected plane of sky map. The magnetic field 
line radius $R$ (curvature $C \equiv 1/R$) will generally also change with 
positions. It is, however, not possible to define a curvature
based on a single isolated polarization segment.
For polarization observations
with a sufficiently large number of segments (as e.g. in Figure \ref{w51_e2}),
a curvature resulting from two adjacent (or more) polarization 
segments can be calculated. 

\begin{figure}
\begin{center}
\includegraphics[scale=0.85]{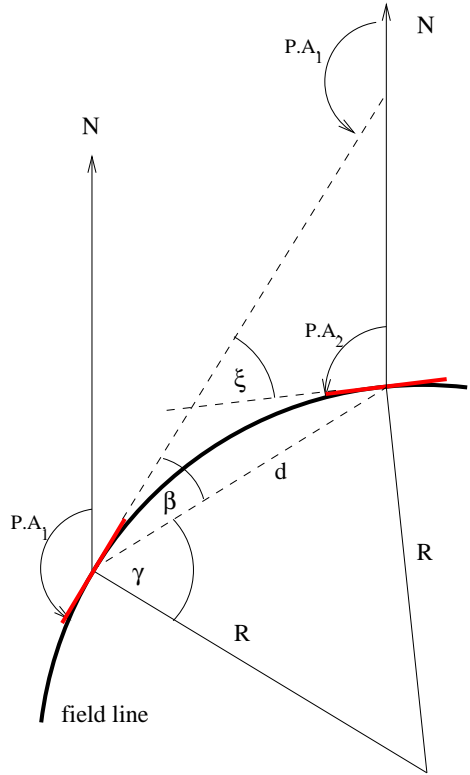}
 \caption{\label{schematic_curvature}
Illustration of the local curvature derivation. Position angles ($P.A.$s), 
separated by an observed distance $d$, are
measured counter-clockwise with respect to north (N). The 2 magnetic field
segments (solid red lines) are assumed to be tangential to a connecting 
field line (thick solid black line) with radius $R$. 
The angle $\xi$ measures the difference
in between 2 $P.A.$s, $\xi=|P.A._{1}-P.A._{2}| \equiv \Delta P.A.$.
$R$ is read from the isosceles triangle with $\cos\gamma = \frac{d/2}{R}$, 
where $\gamma=\frac{\pi}{2}-\beta = \frac{1}{2}(\pi-\Delta P.A.)$. 
This then leads to Equation (\ref{C}).
Note that the derivation uses the measured relative quantities $d$ and 
$\Delta P.A.$. Technically, this allows to detach the magnetic field segments from 
their absolute positions, by keeping their relative separation and 
$\Delta P.A.$, and ensures that a connecting field line can be constructed.
Strictly speaking, even this definition of curvature is not exactly local
because it involves at least 2 segments separated by a distance $d$.}
\end{center}
\end{figure}

Two neighboring magnetic field segments, separated by distance
$d$, are assumed to be tangential to a field line of radius $R$. 
The difference 
in their position angles, $\Delta PA$, is a measure of how much 
the direction of the field line has changed over $d$. The curvature $C$ can then 
be written as (Figure \ref{schematic_curvature}):
\begin{equation}
C\equiv \frac{1}{R}=\frac{2}{d} 
                    \cos\left(\frac{1}{2}\left[\pi-\Delta PA\right]\right) \label{C}.
\end{equation}
With this definition of a local curvature, the calculation of 
the field 
strength $B$ in Equation (\ref{B}) is no longer limited 
to an average field curvature, but can be precise to any 
locally varying field structure.

\begin{figure}
\begin{center}
\includegraphics[scale=0.7]{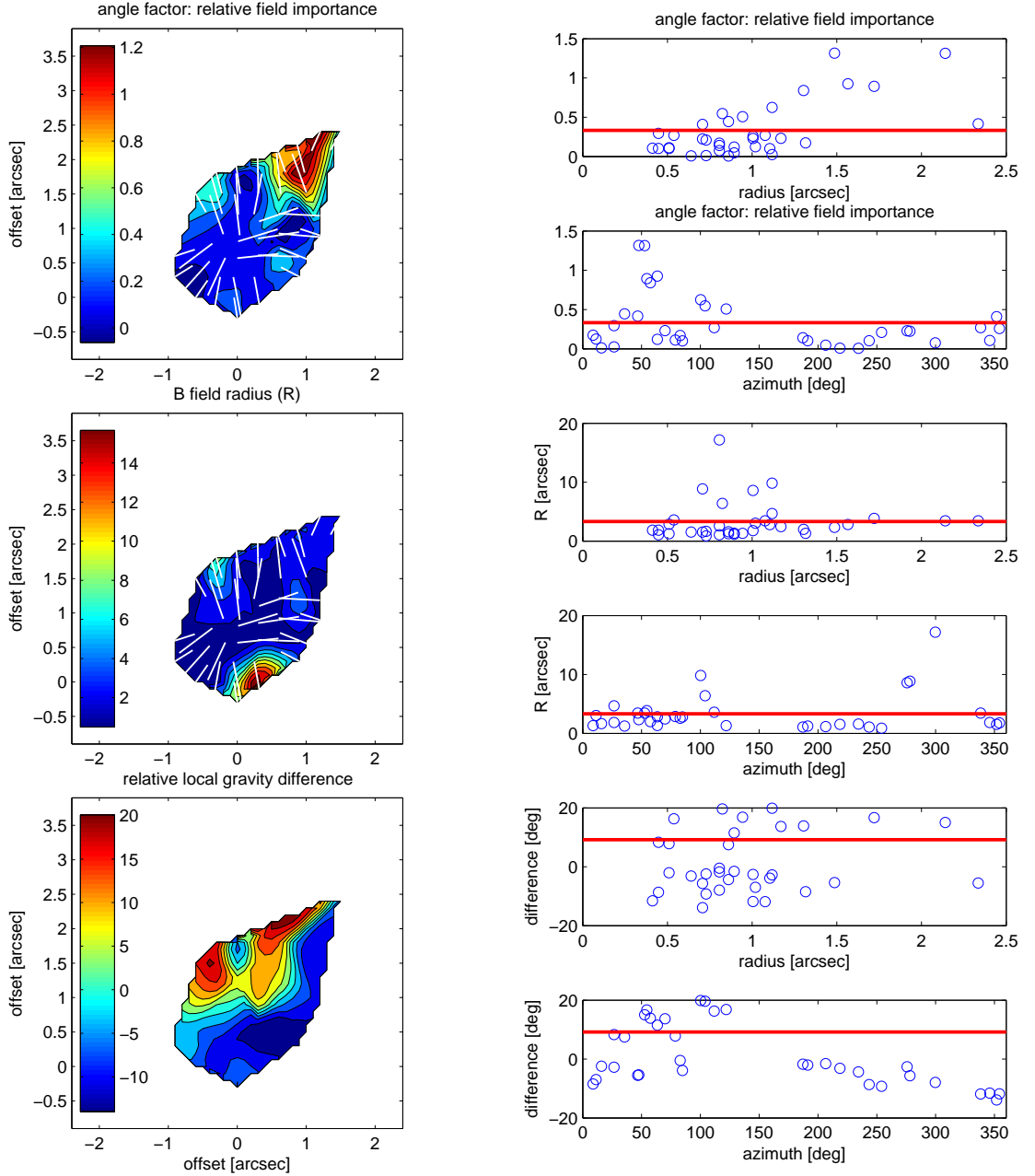}
 \caption{\label{local_factors} \scriptsize 
Position dependence of parameters in Equation (\ref{B}).
Magnetic field segments are shown in white.
Top left panel: the combined factor $\frac{\sin\psi}{\sin\alpha}=\Sigma_B$, 
which quantifies the significance of the magnetic field force compared
to the other forces (Section \ref{local_field_importance}).
$\psi$ is measured with respect to a spherically symmetrical gravitational potential.
Middle left panel: the local magnetic field radius calculated from the method in 
Section \ref{local_curvature}. Units are in arcsec.
Bottom left panel: the difference in the direction of the gravitational pull between
a spherically symmetrical potential and a local gravity direction as defined in 
Section \ref{local_gravity}. Units are in degree. The difference is shown relative
to the symmetrical assumption, revealing a symmetry axis along the southeast- northwest
direction.
The right panels illustrate the tendencies for $\Sigma_B$, $B$ field radius
and difference in gravity direction (from top to bottom) 
as a function of radius (distance from emission peak) and azimuth angle
(measured counter-clockwise from west).
Mean values averaged over the entire maps are shown with the red solid lines.
For the local gravity difference the mean is calculated from the absolute differences.
}
\end{center}
\end{figure}

We remark that the above definition for $C$ is not necessarily 
unique. In particular, determining which segments connect to a field 
line can be non-trivial. Additionally, for some neighboring segments, as it can 
also be seen in Figure \ref{w51_e2}, it is obvious that they can 
not be both tangential to the same field line. The curvature as 
defined in Equation (\ref{C}) is therefore rather a local averaged
curvature. In this view, a local curvature can not only be defined
with two segments which are roughly lining up, but also with two
segments side by side. The local field radius map in Figure 
\ref{local_factors}, middle left panel, 
is calculated by averaging the local curvatures between two 
adjacent segments over all available closest neighbors. A 
closest neighbor is defined through the grid spacing of the map.
Depending on the resolution and the detection of a polarization
structure, the definition of $C$ can possibly be adjusted or refined. 

For W51 e2, the local field radius derived from Equation (\ref{C})
is in the range between $\sim 1\arcsec$ and  $\sim 17\arcsec$ 
(Figure \ref{local_factors}, middle panels). Whereas most areas in the map show rather 
uniform patches with $R\sim 1 \arcsec - 3 \arcsec$, regions in the north-east
and south-west around the accretion plane show larger radii. This
is possibly a consequence of the accretion flow which tends to align
field lines until some limiting inner radius where the field will bend.
Yet more likely, this is a boundary effect because only two almost parallel segments are 
available here for averaging. This leads to unphysically large radii.
The local radius averaged over the map 
is $<R>\approx 3\farcs 3$. Discarding the extreme values gives 
$<R>\approx 1\arcsec$.

\subsection{Local Gravity Direction}  \label{local_gravity}

\begin{figure}
\begin{center}
\includegraphics[scale=0.8]{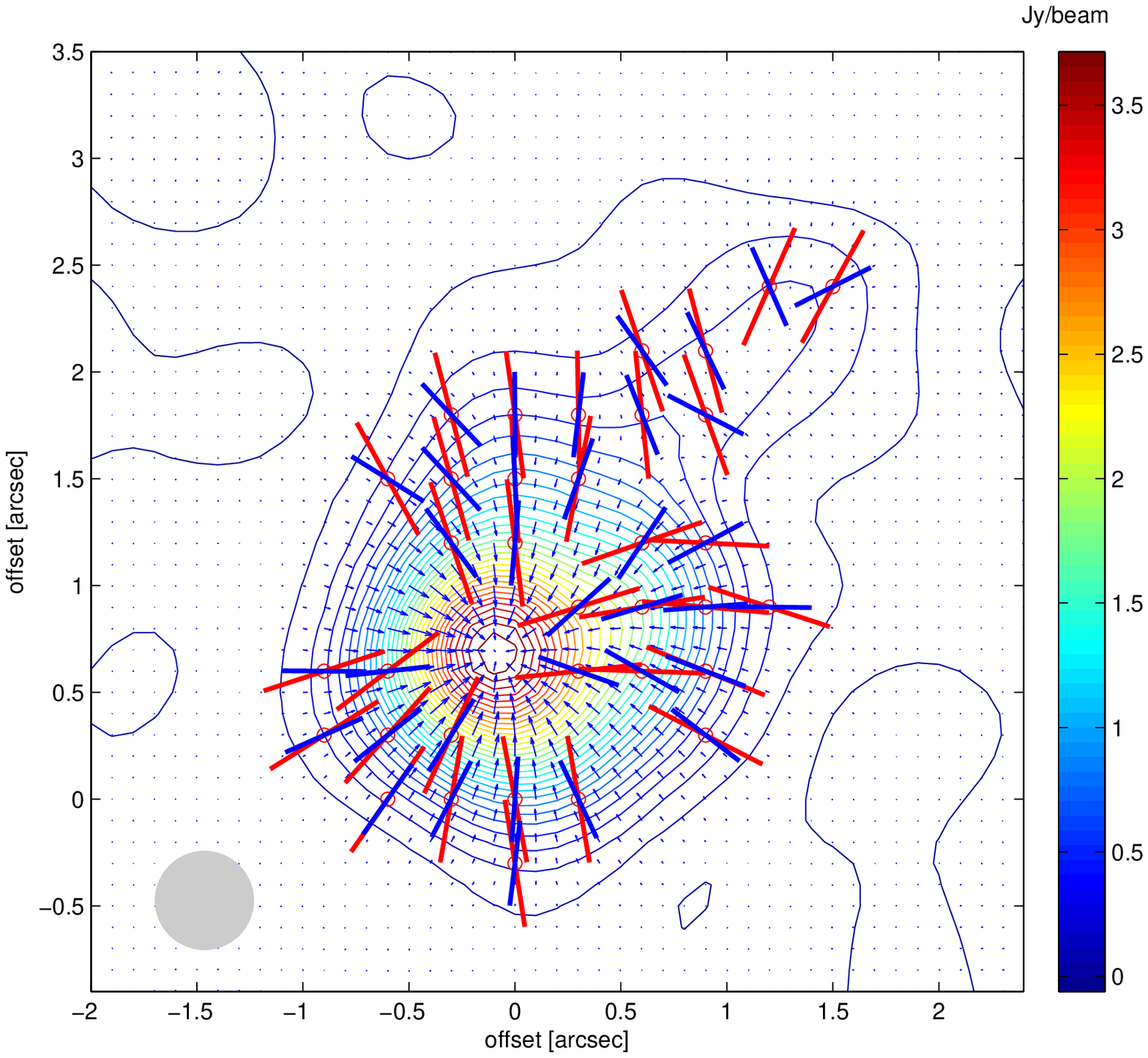}
 \caption{\label{w51_e2}
The collapsing core W51 e2 \citep{tang09}.
Contours show the Stokes $I$ dust continuum at 0.87~mm with 
a synthesized beam resolution of about $0\farcs7$ 
(gray ellipse in the lower left corner) with natural weighting. 
Colors correspond to the color
wedge on the right hand side with units in Jy/beam. Contours are
linearly spaced in steps of 130 mJy/beam.
Overlaid are the magnetic field segments (thick red segments) at the 
locations where polarized 
emission was detected. Magnetic field segments are plotted by rotating the 
polarization segments by 90$^{\circ}$. Only polarization data with a flux 
above $3\sigma_{I_p}$, the rms noise of the polarized intensity, are included.
The blue vector field displays the gradient directions of the dust continuum
emission, with most vectors pointing toward the emission peak.
Highlighted are those emission intensity gradient directions 
(thick blue segments) at 
the locations of the magnetic field segments.
The length of the segments is arbitrary, and for visual guidance only.
The axes offset positions (in arcsec) are with respect to the original 
phase center of the observation at Right Ascension (J2000)=
$19^{\rm h} 23^{\rm m} 43^{\rm s}.95$,
Declination (J2000)=$14^{\circ} 30\arcmin 34\farcs00$
}
\end{center}
\end{figure}

\begin{figure}
\begin{center}
\includegraphics[scale=0.6]{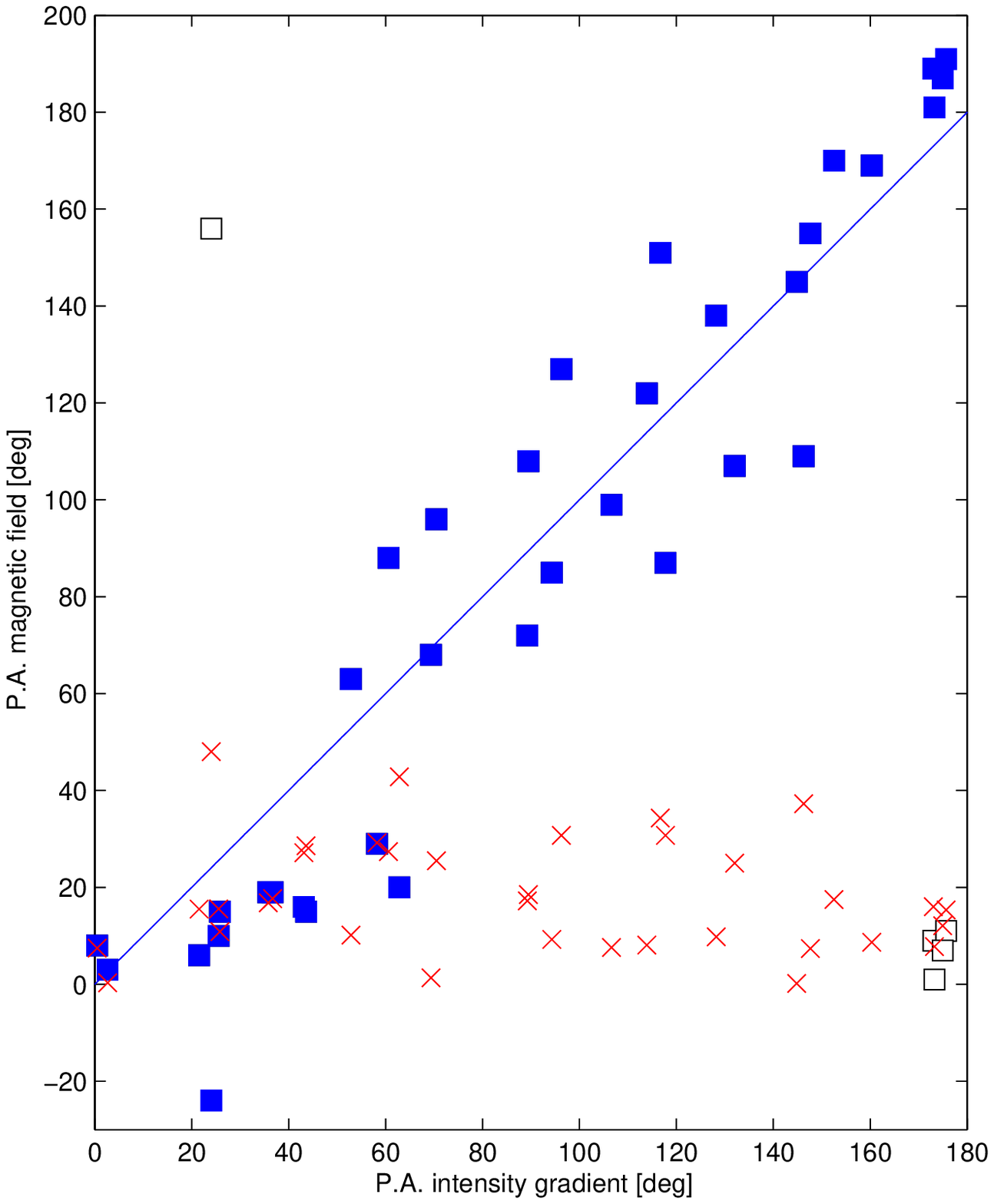}
\includegraphics[scale=0.65]{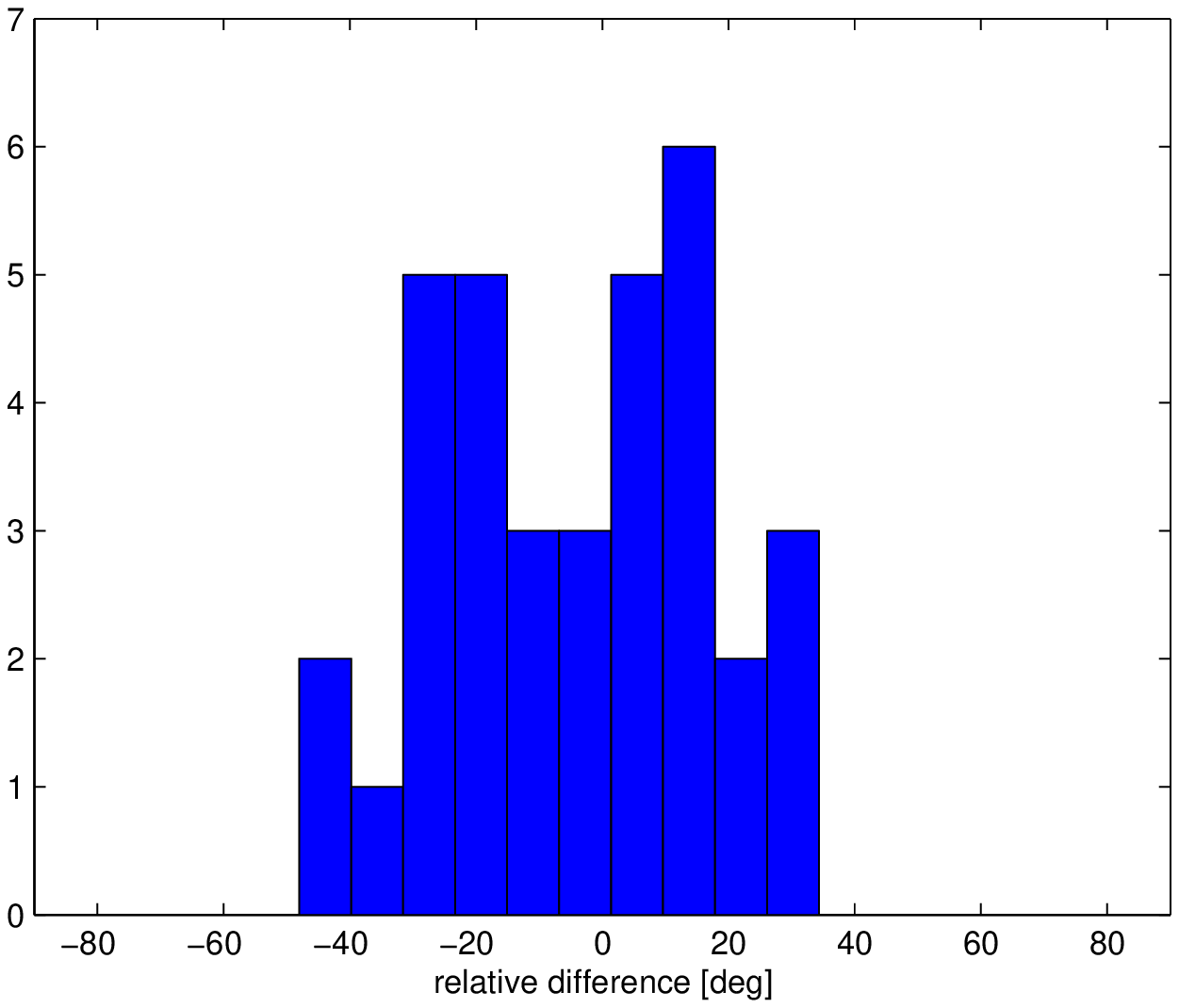}
 \caption{\label{correlation_b_i_grad}\footnotesize
Top panel: the correlation (blue filled squares) between the intensity gradient $P.A.$s 
(blue segments in Figure \ref{w51_e2}) and the 
magnetic field $P.A.$s (red segments in Figure \ref{w51_e2}). 
The black empty squares belong to pairs with $P.A.$s close to $P.A.=0$,
one $P.A.$ being slightly rotated to the left and the other one to the right 
hand side of the vertical. In order to properly display their
correlations, the magnetic field $P.A.$ is re-defined beyond the 
0 to 180$^{\circ}$ range for these cases (blue filled squares above 
180$^{\circ}$ and below 0$^{\circ}$).
For visual guidance added is the straight blue line, 
representing a perfect correlation.
Also shown are the absolute differences ($\le 90^{\circ}$)  
between the $P.A.$s for each pair (red crosses). Both  $P.A.$s are defined
counter-clockwise starting from north.  
Bottom panel: the bimodal distribution of the relative differences 
(in the range between $-90^{\circ}$ to $+90^{\circ}$) between the intensity
gradient and the magnetic field $P.A.$s, reflecting a systematic clockwise and 
counter-clockwise rotation of the field with respect to the intensity gradient. 
}
\end{center}
\end{figure}

\begin{figure}
\begin{center}
\includegraphics[scale=0.85]{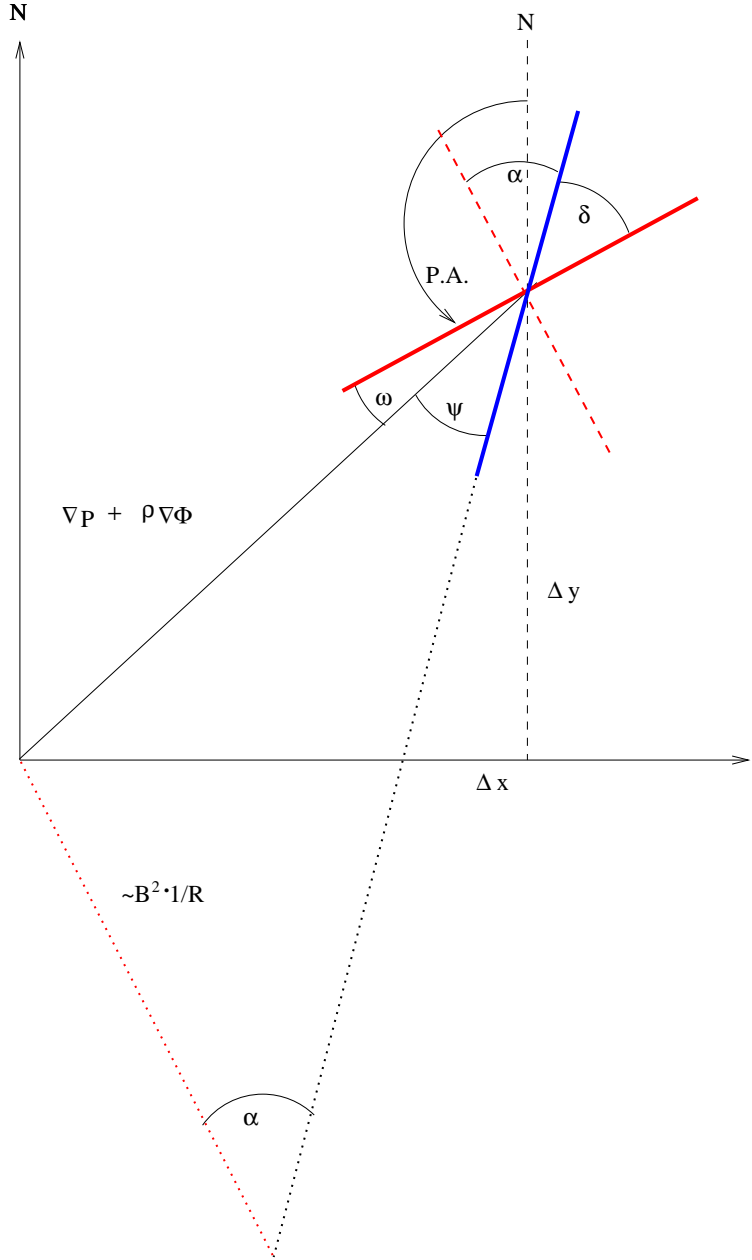}
 \caption{\label{schematic_b}\footnotesize
Illustration of the Equation (\ref{B}) with the relevant angles 
in order to solve for the magnetic field strength $B$.
A pair of magnetic field (red solid line) and emission intensity gradient
segments (blue solid line) are shown at a location $(\Delta x, \Delta y)$ 
with respect to a common reference center.
The originally measured polarization 
orientation
(rotated by 90$^{\circ}$ 
with respect to the field 
orientation)
is indicated as a red dashed line.
Position angles are measured counter-clockwise with respect to north (N).
$\delta \le \pi/2$ measures the angle in between the intensity gradient 
and the field 
orientation.
$\alpha$ is its complement to $\pi/2$.
$(\nabla P+\rho\nabla\phi)$ are assumed to be known in direction and 
strength. For simplicity, as suggested from Figure \ref{w51_e2}
for the case of W51 e2, the emission peak is assumed as the gravity
center, and thus defines direction and strength. 
Here, it also serves as the reference center.
$\psi$ denotes the deviation of the intensity gradient
from the direction of $(\nabla P+\rho\nabla\phi)$. The corresponding 
deviation for the magnetic field is measured with the angle $\omega$.
The force triangle -- $(\nabla P+\rho\nabla \phi)$ pointing toward the 
gravitating center, the resulting intensity gradient 
orientation
(left hand
side in Equation (\ref{mhd_2})) and the direction of the restoring magnetic 
field tension (normal to the field segment) -- can then be closed by 
intercepting the red dotted 
and black dotted lines, where the former one is shifted parallel to 
the red dashed line.
}
\end{center}
\end{figure}

\begin{figure}
\begin{center}
\includegraphics[scale=0.6]{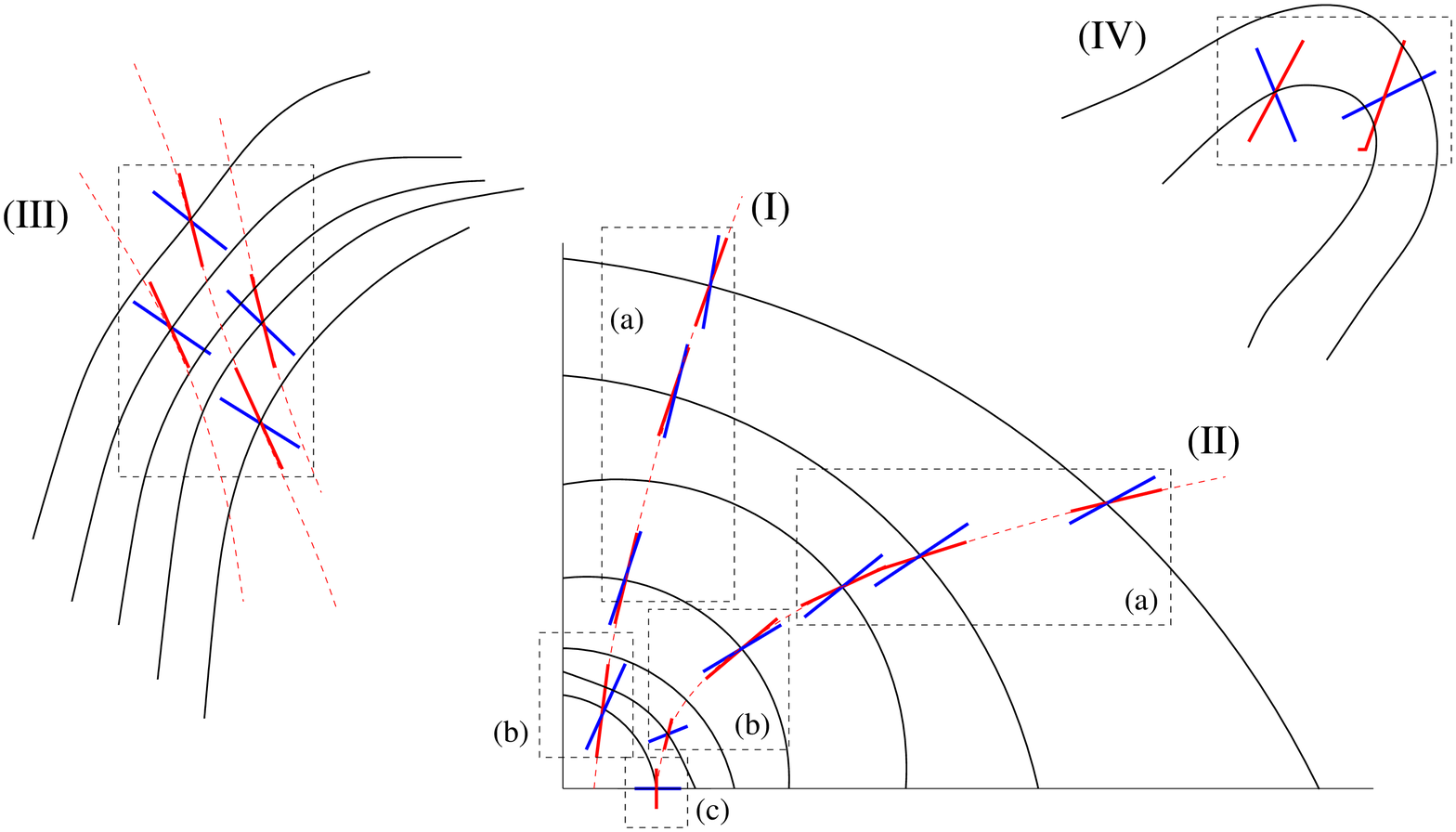}
 \caption{\label{qualitative}
Qualitative analysis (Section \ref{analysis}): Illustration of the expected 
features of the method. The black solid lines are intensity emission contours.
Blue and red segments are intensity gradient and magnetic field 
orientations,
respectively, as in the Figures \ref{w51_e2} and \ref{schematic_b}. 
The labeling refers to the 
cases discussed in Section \ref{analysis} and listed in Table \ref{table_analysis}.
All cases have similar equivalents in the W51 e2 map in Figure \ref{w51_e2}.}
\end{center}
\end{figure}

Unless the gas distribution in a molecular cloud is (close to)
perfectly symmetrical in azimuth around an emission peak, $\nabla \phi$ will differ
in direction and strength from a simple radius-only dependence in the
gravitational potential. Thus, for an arbitrary mass distribution, the 
direction and strength of the local gravitational pull,  
$\rho\frac{\partial\phi}{\partial s_{\phi}}\mathbf{e}_{s_{\phi}}$
in Equation (\ref{mhd_2}), can not any longer simply be constructed
with the direction and distance to the reference center (emission 
peak in Figure \ref{w51_e2}). Instead, at any position in a map, the 
distribution of all the surrounding mass has to be taken into account
to calculate the resulting local gravitational force. Since the 
method in Section \ref{method} is based on measuring angles between
various force directions (including gravity) at all positions in a 
map, any substantial local deviations from a simple global gravity
model can be relevant. 

For a discrete mass distribution ($m_i$), the resulting total potential
$\phi$ at a location $\mathbf{r}$ is: $\phi(\mathbf{r})=-G\sum^n_{i=1}
\frac{m_i}{|\mathbf{r}-\mathbf{r}_i|}$, where $G$ is the gravitational 
constant. We assume that a detected dust emission (like in Figure \ref{w51_e2})
is proportional to the gravitating mass, and that it further traces 
reasonably well the overall gravitating potential. Additionally, it 
is assumed that the dust temperature is roughly constant over the emission
map. In this case, there is no weighting for $m_i$ when linking it to 
the integrated dust emission $I_i$. The local gravity force (per unit mass), 
$\mathbf{g}(\mathbf{r})=-\nabla \phi (\mathbf{r})$ is then:
\begin{equation}
\mathbf{g}(\mathbf{r})=G\sum^n_{i=1}\frac{I_i\cdot f_i}{|\mathbf{r}-\mathbf{r}_i|^2}
                       \cdot \mathbf{e}_i,   \label{G}
\end{equation}
where $f_i$ is a conversion factor linking dust emission and total 
mass, and $\mathbf{e}_i$ is the direction between each detected 
position with dust emission ($\mathbf{r}_i$) and the local position 
($\mathbf{r}$) where the gravity force is evaluated.

We remark that, if $f_i\equiv f$ over the entire map, the local 
gravity direction can be derived without prior knowledge of $f$. 
For a typical observation, the mass (intensity) distribution is 
obtained by pixelizing a map with the beam resolution. Figure 
\ref{local_factors}, bottom left panel, shows the deviations from
an azimuthally symmetrical potential for W51 e2 when calculating the local
gravity directions following Equation (\ref{G}) for $f_i\equiv f$.
This analysis shows that, although W51 e2 (Figure \ref{w51_e2}) 
appears to be a rather
spherically symmetrical collapsing core, the deviations are 
systematically positive or negative on one or the other side of 
the hourglass axis. Minimum deviations are roughly along the 
south-east north-west direction which is consistent with the 
hourglass axis. 
The average 
absolute
deviation is 
$\approx 9^{\circ}$. 
As further discussed in the Sections \ref{dynamic_b} and \ref{b_gravity}, 
already this simple symmetry analysis points toward a systematic influence
of the magnetic field.

These deviations provide a correction for the angle $\psi$ (angle 
between the gravity direction and the intensity gradient, Figure 
\ref{schematic_b}), which originally is calculated assuming an 
azimuthally symmetrical potential (Figure \ref{angles}, top left panel).
Consequently, also the magnetic field strength $B$ in Equation (\ref{B})
needs to be recalculated. The angle $\alpha$, as defined relative
to the intensity gradient, remains unchanged with this local gravity correction.

We finally note that further refinement of the local gravity correction
introduced here is possible. Local over-densities -- which follow the flow 
of the global dynamics -- can eventually locally collapse once they pass
the Jeans mass. The north-west extension in Figure \ref{w51_e2} is 
possibly such a case. Using Equation (\ref{G}), however, the resulting
local gravity direction will always be dominated by the main mass (emission).
Consequently, the angle $\psi$ might be over- or underestimated.
This defect can be corrected by identifying local collapse centers
with a local maxima search and a suitable threshold criterion.

\section{Results}  \label{results}

This section applies the method developed in Section \ref{method}
to the case of W51 e2.
Starting from the expression for the magnetic field strength
$B$ in Equation (\ref{B}), results are first presented for separated 
factors and then for the final field strength.
Whereas in the original observed map in Figure \ref{w51_e2} the magnetic
field is gridded to half of the synthesized beam resolution, the maps 
in the following are interpolated for an enhanced visual impression.
This can emphasize features along the field lines where polarization
is detected at equally separated spacings. On the other hand, it might
lead to non-physical values in depolarization zones along 
the south-west north-east direction or in the center, where no data are available.

\subsection{Angle Factors}

\subsubsection{$\sin \psi$ -- map}

\begin{figure}
\begin{center}
\includegraphics[scale=0.65]{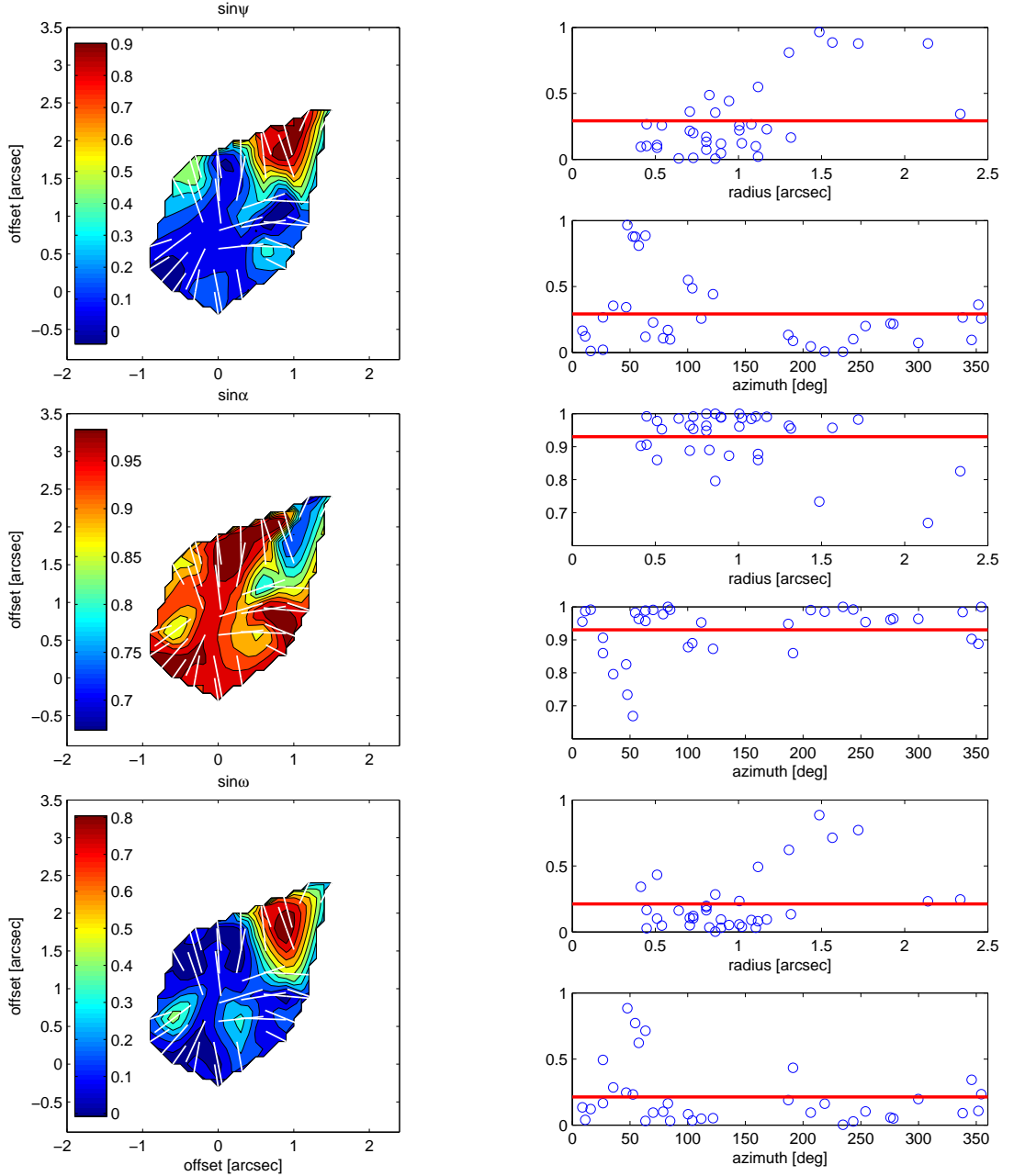}
 \caption{\label{angles} \footnotesize
Relevant angles for the derivation of the magnetic field strength, calculated
for W51 e2 (Figure \ref{w51_e2}).
Magnetic field segments are shown in white.
Top left panel: $\sin\psi$-map, where $\psi$ $(\le\pi/2)$ (Figure \ref{schematic_b})
is the difference between the  
orientations
of the intensity gradient and gravity.
Middle left panel: $\sin\alpha$-map. $\pi/2-\alpha$ $(\le\pi/2)$ measures the deviation
between the magnetic field and the intensity gradient
orientations.
Bottom left panel: $\sin\omega$-map, where $\omega$ is the difference between the 
magnetic field and gravity 
orientations.
The right panels show the corresponding tendencies for each angle 
as a function of radius (distance from emission peak) and azimuth angle
(measured counter-clockwise from west).
Mean values averaged over the entire maps are shown with the red solid lines.
}
\end{center}
\end{figure}

The angle $\psi$ measures the difference between the
orientations
of 
the intensity gradient and gravity. Small values in $\psi$ (or
$\sin\psi$) therefore indicate that a change in the local intensity
structure is closely following global gravity. 
The top panels in Figure \ref{angles} show a map
of $\sin\psi$ together with its trends as a function of radius 
and azimuth. The peak emission (Figure \ref{w51_e2}) is again assumed as 
the reference center, and the gravity direction is simply taken as 
the radial direction from the center to any measured intensity 
location.
Small values of $\sin\psi$ - in the range
between 0 and 0.2 - are found for most of the magnetic field segments
pointing toward the center. An X-pattern is revealed made up from 
the majority of the segments. There is a trend of larger $\sin\psi$
values in the north-east and south-west where likely accretion is 
ongoing. The magnetic field segments in the far north show $\sin\psi$
values close to one. Here, the intensity gradient directions are 
very different (by up to 90$^{\circ}$) from the global gravity directions.
This northern patch, therefore, seems to be decoupled from the 
main collapsing core (Section \ref{analysis}, case (IV)).
$\sin\psi$ averaged over the entire map
gives $<\sin\psi>\approx 0.29$ ($<\psi>\approx 20^{\circ}$). 
The average decreases to about 0.19 ($<\psi>\approx 12^{\circ}$)
if the six polarization segments in the far north are ignored.
A hint of an increase in $\sin\psi$ with radius is apparent in 
the top right panel in 
Figure \ref{angles}. Again, the most northern patch dominates values
beyond a distance of $\sim 1.5\arcsec$. No clear structure in azimuth 
is revealed.

Since $\psi$ measures a deviation from gravity, mass distributions
differing  from an azimuthal symmetry can introduce corrections
to the local gravity direction (Section \ref{local_gravity}).
The top left panel 
in Figure \ref{angles_local_gravity} displays the $\sin \psi$ -- map 
calculated from the generalized 
gravitational potential in Equation (\ref{G}).  
The overall patterns are still similar to the uncorrected
$\sin \psi$ -- map, but with smaller values in the accretion areas.
As a result, the average is reduced to $<\sin\psi>\approx 0.19$
($<\psi>\approx 12^{\circ}$)
and to about 0.09 ($<\psi>\approx 6^{\circ}$) when discarding 
the most northern patch.

\subsubsection{$\sin\alpha$ -- map}

The difference in
orientations
between the intensity gradient and the 
magnetic field is denoted with $\delta$ (Figure \ref{schematic_b}). The angle
$\alpha$ is its complement to 90$^{\circ}$. Or, in other words, $\alpha$ 
is the difference in 
orientations
between a measured (projected) 
polarization and the intensity gradient. Values of $\alpha$ close to 
its maximum of 90$^{\circ}$ ($\sin\alpha\approx 1$), indicate that
the projected magnetic field 
orientation
and the intensity gradient
are closely aligned. 
This again reflects the initially presented correlation in Section \ref{motivation}.
The middle panels in Figure \ref{angles} show
a map of $\sin\alpha$ with its radial and azimuthal dependence. 
Around the collapsing core, the values of $\sin\alpha$ are 
between 0.85 and 1. They decrease to about 0.7 in the most northern 
area. The intensity and magnetic field 
orientations
are therefore 
very similar over most of the map, with an average value 
$<\sin\alpha>\approx 0.93$ ($<\alpha>\approx 72^{\circ}$). 
Leaving out the most northern part gives
$<\sin\alpha>\approx 0.95$ ($<\alpha>\approx 74^{\circ}$). 
Possibly there is a slight trend of 
decrease in $\sin\alpha$ with radius. No clear structures appear
in azimuth. $\alpha$ is not affected by local gravity corrections
as outlined in Section \ref{local_gravity}.

\subsection{Magnetic Field Force Compared to Other Forces}
\label{result_mf_force}

The angle factor $\left(\frac{\sin\psi}{\sin\alpha}\right)_{local}=
\left(\frac{F_B}{|F_G+F_P|}\right)_{local}\equiv \Sigma_B$
measures in a model-independent way the local impact of the magnetic field
force (Section \ref{local_field_importance}). The top panels in Figure 
\ref{local_factors} illustrate the results.
Values for the relative field 
significance, $\Sigma_B$, 
range from $\sim 0.1$ to $\sim 1.3$, with an average over the entire map 
$<\Sigma_B>\approx 0.33$. Thus, the 
magnetic field can balance only about one third of the gravitational force. 
Consequently, it can not prevent the core from any further collapse.
We stress that this result is obtained without any assumption about the 
mass involved. The top panels in Figure \ref{local_factors} further 
show that the impact of the magnetic field changes as a function of 
position. Generally, the field influence seems to become weaker
toward the center where gravity likely becomes more and more dominant.
This result is further explored in \citet{koch11}.

\subsection{Magnetic Field Strength Maps} \label{result_field_strength}

Maps for the magnetic field strength $B$ are presented in this section for 
two scenarios: Firstly, only the newly introduced force ratio $\Sigma_B$ is 
taken into account, leaving the other parameters in Equation (\ref{B}) constant.
This will illustrate the influence of $\Sigma_B$. Secondly, the field strength
is calculated  by combining $\Sigma_B$ together with an analysis of the density
profile and the resulting gravitational potential.
A first order estimate of the field strength, comparing the 
gravitational force $F_G=\frac{G M_R \rho}{R_G^2}$ and the magnetic field 
tension $F_B=\frac{B^2}{4 \pi R_B}$ is given in \citet{tang09}. Since
W51 e2 is a collapsing core, $F_G>F_B$, an upper limit of the field 
strength, $B<19$~mG, can be derived. 
$M_R (\approx 220~M_{\odot})$ 
refers to the gas mass enclosed within a radius $R_G (\approx 1\arcsec)$.
The radius of a magnetic flux tube is $R_B (\approx 1\arcsec)$. The mass
density $\rho$ is estimated with the gas volume number density 
$n_{H_2} (\approx 2.7 \times 10^7$~cm$^{-3})$. $G$ is the gravitational
constant.
The above value, $B<19$~mG, will serve as a reference for the magnetic field strength 
maps presented in the following.
We remark that Equation (\ref{B}) is consistent with the above estimate,
$B=\sqrt{4\pi R_B G \frac{M_R \rho}{R_G^2}}=\sqrt{4\pi R_B \rho\nabla\phi}$,
when setting $\frac{\sin\psi}{\sin\alpha}=1$ and neglecting the pressure 
gradient $\nabla P$.

\begin{figure}
\begin{center}
\includegraphics[scale=0.65]{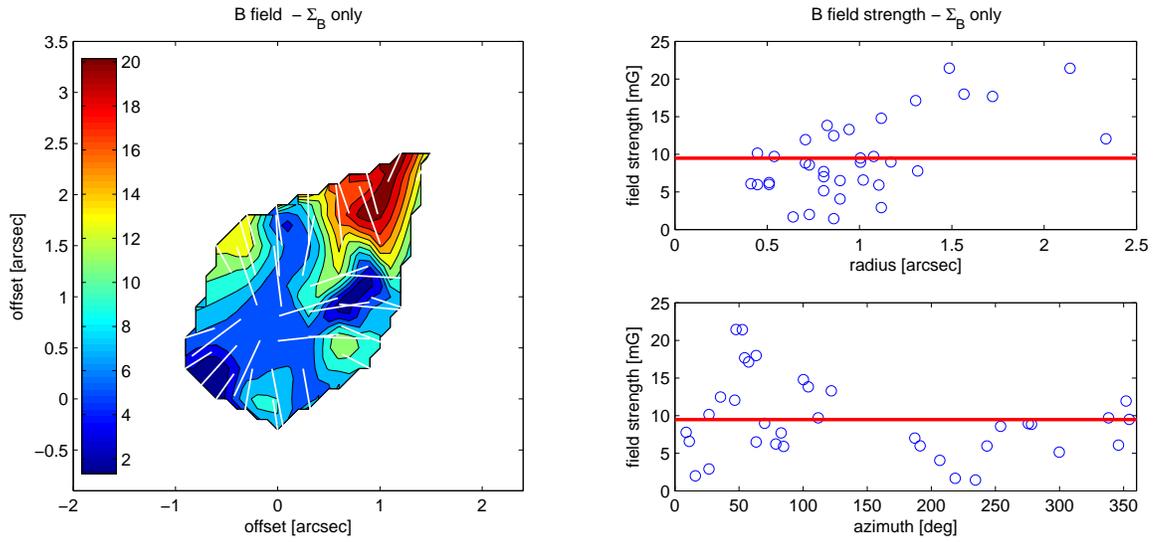} 
 \caption{\label{map_b}
The magnetic field strength (B) derived by taking into account only 
variations in the field-to-gravity force factor
$\frac{\sin\psi}{\sin\alpha}\equiv \Sigma_B$.  Units are in mG. 
All the other parameters in Equation (\ref{B}) are constant over
the map.
Magnetic field segments are shown in white.
The right panels show the field strength tendencies
as a function of radius (distance from emission peak) and azimuth angle
(measured counter-clockwise from west).
Mean values averaged over the entire map are shown with the red solid lines.
}
\end{center}
\end{figure}

In the first scenario, the $B$ map for W51 e2 is calculated by taking into account only
the angles  $\psi$ and $\alpha$. Field radius, potential and density are set
constant as above. Assuming local changes in temperature and density to be
negligible compared to gravity,
the pressure gradient $\nabla P$ is omitted from Equation (\ref{B})
in the following. From the measured ranges in $\psi$ and 
$\alpha$ (Figure \ref{angles}, $\sin\psi\approx 0.1 - 0.9$, $\sin\alpha\approx 0.7-1$),
it is then obvious that mostly $\psi$ will dominate the range and structure
in a magnetic field strength map (Figure \ref{map_b}).
Values of the factor $\frac{\sin\psi}{\sin\alpha}=\Sigma_B$ range from $\sim 0.1$ 
to $\sim 1.3$, with an average over the entire map 
$<\Sigma_B>\approx 0.33$
(Figure \ref{local_factors}, top panels).
The resulting 
patterns are very similar to the ones in the $\sin\psi$-map 
(Figure \ref{angles}, top left panel). Resulting variations in $B$ are 
then from a few mG to $\approx 20$ mG. The map average value is $<B>\approx 10$~mG.
Compared to the reference value above, the correction factor 
$\frac{\sin\psi}{\sin\alpha}$ generally leads to a lower field strength.
Discarding the most northern patch -- where $B$ is likely overestimated
due to the lack of a clear local gravity center (Section \ref{analysis}) --
the value becomes $<B>\approx 8$~mG.
Obviously, when leaving curvature, gravitational potential and density 
constant, the radial profile of the field strength is determined by $\Sigma_B^{1/2}$.
As the field-to-gravity force ratio decreases toward the center, consequently
the field strength also decreases  and its largest values tend to be in the 
outer zones (Figure \ref{map_b}). This result needs to be compared with 
the following complete analysis.

\begin{figure}
\begin{center}
\includegraphics[scale=0.65]{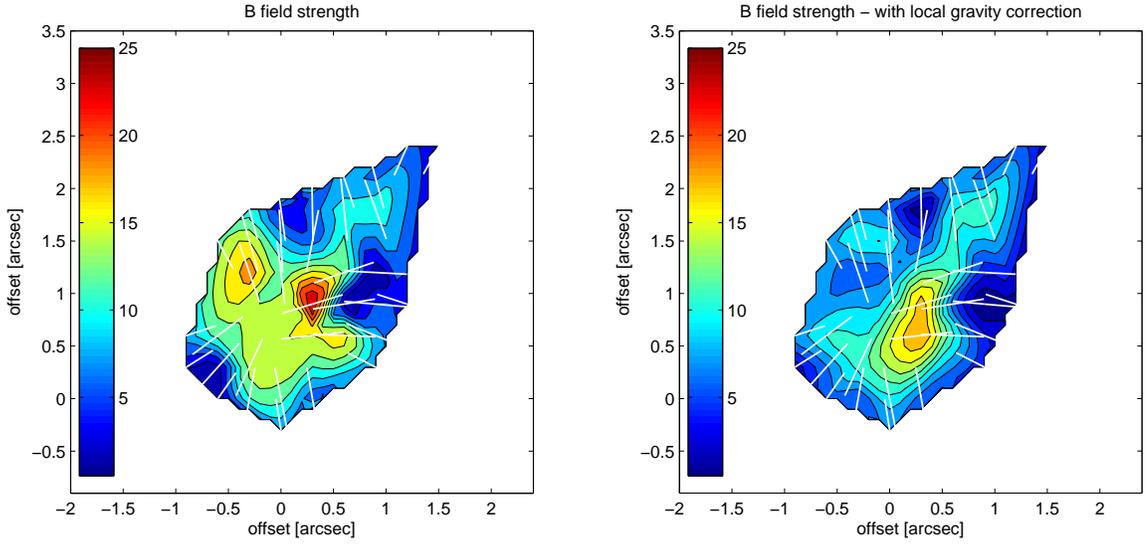}
 \caption{\label{b_map_local}
Final magnetic field strength maps calculated from Equation (\ref{B}). 
Units are in mG. Taken into account are the deprojected profile for 
the density, the resulting gravitational potential gradient profile and 
variations in the force ratio $\Sigma_B$ (left panel). 
In the right panel, corrections due to the local gravity direction
are additionally considered. 
Mean field strengths are about 9.4~mG and 7.7~mG for the left and 
right map, respectively.  The field radius is set constant.
Magnetic field segments are shown in white.
Typical measurement uncertainties lead to an average systematic error in the 
field strength of $<\Delta B>_{sys}\approx\pm 1$~mG.
}
\end{center}
\end{figure}

\begin{figure}
\begin{center}
\includegraphics[scale=0.65]{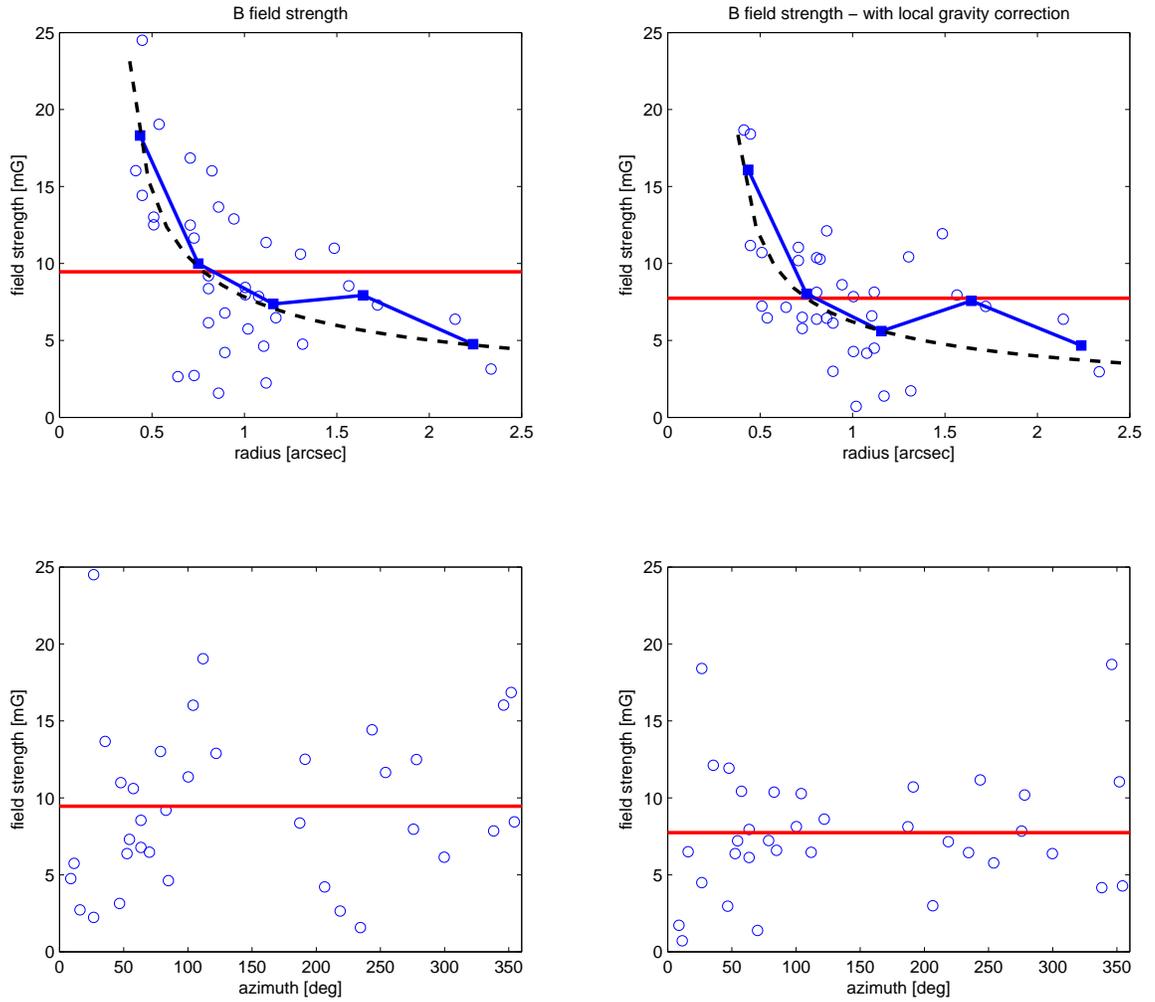}
 \caption{\label{b_map_local_profile}
Radial and azimuthal trends for the magnetic field strength maps 
presented in Figure \ref{b_map_local}. Radial profiles $\sim r^{-1/2}$
are overlaid for illustration in the top panels.
Mean values averaged over the entire map are shown with the red solid lines.
}
\end{center}
\end{figure}

In the second scenario, the field strength is calculated by additionally taking 
into account the profiles of density and gravitational potential. As discussed
later in Section \ref{projection}, the force ratio $\Sigma_B$ is little or 
not at all affected by projection effects. Therefore, we proceed to reconstruct
3-dimensional radial profiles assuming spherical symmetry. For that purpose, 
the azimuthally averaged dust emission profile from Figure \ref{w51_e2} is
deprojected with an Abel integral. The resulting 3-dimensional radial profile
($r$) is well described with a functional form $f(r) \sim \frac{1}{(r+0.35)^3}$, 
whereas the initial projected map ($R$) is well fit with a profile 
$\sim \frac{1}{(R+0.25)^2}$. The fitting is performed on profiles with values
binned to half of the synthesized beam resolution. Assuming the total 
mass of the e2 core to be distributed similar to its dust component, the 
cumulative mass $m(r) \sim \int_0^r f(r^{\prime})\, 4\pi r^{\prime 2}\,dr^{\prime}$
and the resulting gradient of the potential $\sim \frac{m(r)}{r^2}$ can 
be derived. Combining all together (Equation(\ref{B})), the magnetic field 
strength profile scales as 
$B(r)\sim \left(\Sigma_B(r)\cdot f(r) \cdot \frac{m(r)}{r^2}\right)^{1/2}$.
Whereas the dust emission map allows us to derive this functional form, its 
absolute normalization remains to be determined. Adopting $<\Sigma_B>\approx 0.33$
(Section \ref{result_mf_force}) -- i.e. the observed averaged magnetic 
field strength can balance only about one third of the gravitating mass
$M_R (\approx 220~M_{\odot})$ -- we normalize $B$ so that 
$<B^2>=<\Sigma_B>\cdot (19~mG)^2$, where the latter field strength is equivalent
to $220~M_{\odot}$ \citep{tang09}. This gives $<B>\approx 9.4~$mG. 
The final field strength map and its radial and azimuthal trends are shown in 
the left panels in the 
Figures \ref{b_map_local} and \ref{b_map_local_profile}. 
An increase in field strength with smaller radius becomes apparent.
A profile $\sim r^{-1/2}$ is overlaid for illustration. A hint of an increase
in field strength around $r\approx 1.5\arcsec$ possibly points toward the 
forming core in the northwest. This is found by still assuming a dominating 
gravity center at the emission peak. It is obvious that the first map 
(Figure \ref{map_b}) can not represent the correct field strength, but merely
reflects the relative field significance. Only by adding profiles for the density
and the gravitational potential can the magnetic field profile correctly
be derived. Density and gravitational potential only (setting $\Sigma_B \equiv 1$)
would lead to a field profile $\sim r^{-1.7}$. Taking into account the 
decreasing field-to-gravity force ratio with radius, leads to the shallower
scaling $B(r)\sim r^{-1/2}$. With the density profile, $\rho\sim r^{-3}$, 
this connects field and density as $B(r) \sim \rho^{1/6}$.
Whereas the dust emission shows little variations in azimuth (close to 
spherical symmetry), the local position-dependent measurement of $\Sigma_B$
shows some variations. Consequently, features in the magnetic field strength
map start to be revealed, with possibly regions of increased strength in the 
accretion zones in the north-east and the south-west directions. In order to 
trace the magnetic field to the very center, higher resolution observations
will be needed. 

We remark that the projected dust emission can also reasonably well, although 
with larger residuals, be fit with a $\sim R^{-1}$ profile. The deprojected 
profile $f(r) \sim \frac{1}{(r+0.25)^2}$ then gives a magnetic field profile
$B(r) \sim r^{-0.3}$ and $B(r) \sim \rho^{3/20}$ (not shown). This leads to about 15\% 
lower values in the field strength in the center with a similar mean field
strength $<B>\approx 9.8~$mG. All the other features are left unchanged. In 
the above final field strength map we have neglected the variations in the 
field radius and assumed it to be constant (Figure \ref{local_factors}, top panels).
This ignores two zones where large field radii are likely a consequence
of boundary effects (Section \ref{local_curvature}).

Finally, the influence of the local gravity correction is illustrated in the 
right panels in the Figures \ref{b_map_local} and \ref{b_map_local_profile}.
Generally, features tend to be smoother. The field strengths in the 
accretion zones are less pronounced and a single center becomes apparent. 
The average strength is about 7.7~mG. The radial profile of the binned 
values remains close to $B(r)\sim r^{-1/2}$ with maximum values around 19~mG.
Structures in azimuth remain similar although with smaller variations.  

We stress that, although we are making use of the total gravitating mass
$M_R$ -- which sets an upper limit for the field strength and defines its 
order of magnitude -- only the newly derived force ratio $\Sigma_B$ leads to
a proper normalization and opens the possibility to derive a radial profile. 
This has become possible by solving the MHD force equation by the method 
introduced in Section \ref{method_detail}. 

Errors $\Delta B$ in the final $B$ field maps are estimated by propagating
typical measurement uncertainties through Equation (\ref{B}). The individual
polarization $P.A.$ uncertainty of $5^{\circ}$ to $10^{\circ}$ is assumed
for $\Delta \alpha$. A slightly smaller uncertainty, $\sim 3^{\circ}$ to
$5^{\circ}$, can be expected for $\Delta\psi$ because of the averaging
process in the interpolation when calculating the intensity gradient.
This leads to $\Delta B\approx \pm\frac{1}{2B}(4\pi R\rho\nabla\phi)\cdot 0.05$,
where the numerical factor follows from the above uncertainties combined with 
map averaged values for $<\sin\alpha>$ and $<\sin\psi>$. For any of the 
above scenarios and $<B>\approx 9$~mG, we derive an average systematic
error $<\Delta B>_{sys}\approx \pm 1$~mG. 
Errors in the profiles for density and gravitational potential are small
after binning and azimuthal averaging. They are, therefore, neglected here.
Besides this 
systematic error, there is an additional statistical uncertainty in the
field 
orientation
resulting from the turbulent field dispersion. Adopting
$5^{\circ}$ to $10^{\circ}$ for this \citep{koch10}, this contributes to 
the final error budget with $<\Delta B>_{stat}\approx \pm 1$~mG.
We note that all uncertainties in angles are smaller than the mean value
($\approx 20^{\circ}$) of the differences between magnetic field and 
intensity gradient $P.A.$s (Section \ref{motivation}). This further
confirms that the correlation in Figure \ref{correlation_b_i_grad} can 
not only result from systematic and statistical uncertainties.

In conclusion, density and gravitational potential profiles determine
the order of magnitude of the field strength. The local features result 
from the angle factor $\frac{\sin\psi}{\sin\alpha}$.
These features remain robust with typical total errors (systematic and statistical)
of $\Delta B\approx \pm (1-2)$~mG.
The derived field strengths are also consistent with the level of 
magnetic fields typically detected by Zeeman effects on OH masers in 
compact HII regions (a few mG up to $\sim 20$~mG, \citet{fish07}).

\section{Discussion}    \label{discussion}

\subsection{Comparison, Application Regime and Limitations of the Method}
\label{application}

The method presented in Section \ref{method} leads to a local magnetic
field strength. It, nevertheless, uses 
strictly speaking a not exactly local but larger scale
property of the 
field, namely its curvature (Figure \ref{schematic_curvature}). 
The shape of a field line is the result of
various forces acting on the dust particles coupled to the magnetic field.
These can be both long-ranging forces (like gravity) as well as locally
acting perturbations (like turbulence). In order to solve 
Equation (\ref{mhd_momentum}), we need to determine the larger scale 
curvature. The locally turbulent field can here act as a contaminant.
This is in contrast to, e.g. the CF method which precisely uses the field
dispersion around an average field direction, in combination with some 
turbulent velocity information, to derive a magnetic field strength. 
In principle, a curvature can be calculated from a minimum of two segments
(Section \ref{local_curvature}), but the local turbulent field can 
introduce some uncertainty here. Assuming an isotropic turbulent field 
component, a local curvature can be derived by averaging over several
neighboring curvatures. As long as reasonably smooth and connected
polarization patterns are detected, this will reveal the larger scale
field curvature. This is the case for the field morphology in Figure \ref{w51_e2}
where a clear signature of dragged-in field lines is apparent. Higher 
resolution data are then needed to trace the field in the inner most
part of the collapsing cloud where the curvature is expected to change
faster within a small area.

The method further relies on identifying a noticeable effect from the 
various forces in Equation (\ref{mhd_momentum}) which shape the field 
on large scales. When assuming flux freezing, this means that
$B^2/(8\pi)<P_{gas}+\rho\phi+\rho v^2$. In this weak field case of 
flux-freezing, the field lines are forced to move along with the gas
(at least up to some scale where ambipolar diffusion sets in). As a 
consequence, gravity, intensity gradient and field 
orientations
can be
individually determined which is needed to calculate the angles in Equation (\ref{B}).
In the strong field case, $B^2/(8\pi)\gg P_{gas}+\rho\phi+\rho v^2$,
where matter can mostly move only along field lines, the method will
yield $\psi \sim 0$, $\alpha\sim\pi/2$, $R\rightarrow \infty$, and 
therefore $B\rightarrow \infty$.
(see also Section \ref{analysis}.) Unless the pressure gradient $\nabla P$
can be identified, which then allows to measure $\psi\neq 0$, the method
can not further constrain $B$ (see also Table \ref{table_analysis}).
Geometrically, the method is based on closing a force triangle 
(Figure \ref{schematic_b}). Thus, if $\psi\sim 0$ or $\alpha\sim 0$ or if 
both angles are zero at the same time, the triangle is no longer defined 
and the method fails.

In summary, whereas the CF method is based on the field dispersion but 
has to isolate a mean large-scale field direction, the method discussed
here uses the large-scale field morphology shaped by various forces. 
Therefore, the method is more directly applicable to the weak field case. 
A successful extension to the strong field case is possible if pressure
and/or gravitational potential gradients can be localized. Generally, 
coherent patches with smooth changes in field and intensity gradient
directions are of advantage as this leads to a clear identification 
of the field curvature and the 
acting forces. The CF method needs both dust polarization continuum and
molecular line velocity observations. The method here uses the dust continuum
Stokes $I$ and its polarized emission.
Whereas the CF method provides one single statistically averaged value, 
the approach here leads to a map of field strength values.

Finally, we conclude this section by noting that the method here is not limited
to dust continuum polarization. As long as a force triangle can be identified, 
the method is applicable to any measurement which reveals the large-scale
field morphology in combination with gravity and/or pressure forces.
Notably, molecular cloud polarization maps obtained from the Goldreich-Kylafis
(GK) effect, i.e. from spectral line linear polarization \citep{gk81, gk82}, can be 
analyzed in the same way. 
This might open a way to study outflows which are often not observed in dust
continuum due to their relatively small masses.
The only complication possibly rises from the 
difficulty that it can be non-trivial to conclude whether the 
magnetic field is parallel or perpendicular to the
polarization detected from the GK effect. 
In an even broader context,  the method introduced here can also be applied
to field morphologies detected from polarization by synchrotron emission or 
from Faraday rotation measures. The method is, therefore, not limited to 
molecular clouds, but can potentially also be adopted for, e.g. galaxy rotation
measures.

\subsection{Projection Effects} \label{projection}

\begin{figure}
\begin{center}
\includegraphics[scale=0.85]{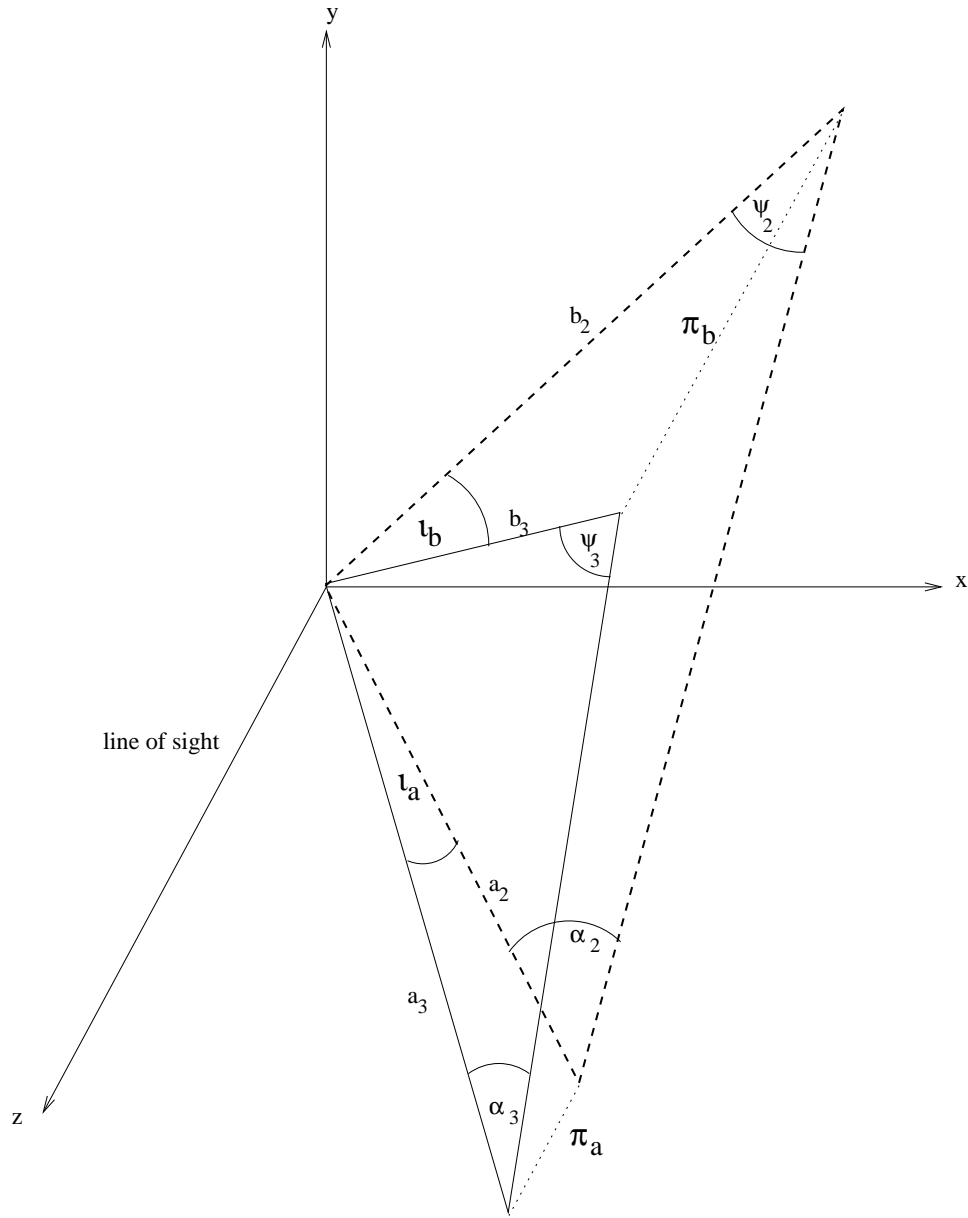}
 \caption{\label{schematic_3d}\footnotesize
Projected and deprojected triangles for the method presented in Section \ref{method}.
Lower indices {\it '3'} or {\it '2'} refer to 3-dimensional or 2-dimensional projected
quantities. The projected triangle (dashed lines) in the $xy$-plane is deprojected 
along $\pi_b$ and $\pi_a$ (parallel to the line of sight) to form a 3D triangle 
(solid lines). The angles $\psi_2$ 
and $\alpha_2$ are identical to $\psi$ and $\alpha$ in Figure \ref{schematic_b}.
$\iota_b$ and $\iota_a$ are the inclination angles of the gravitational pull and the 
magnetic field tension, respectively.
}
\end{center}
\end{figure}

The results in Section \ref{results} were derived assuming the plane of sky projected
geometry in Figure \ref{schematic_b}. Here, we address the question of how an original 3-dimensional
configuration affects the parameters in Equation (\ref{B}) to calculate the magnetic field
strength. The schematic in Figure \ref{schematic_3d} illustrates the connection. We note, that 
like in the 2-dimensional (2D) case, gravitational pull (and pressure gradients), magnetic 
field tension and the resulting intensity gradient have to form a closed triangle. The 
projected 2D triangle can generally be deprojected with 2 different inclination angles:
$\iota_a$ and $\iota_b$, corresponding to the inclinations of the deprojected
orientations
of the magnetic field tension and the gravitational pull, respectively. 

\begin{figure}
\begin{center}
\includegraphics[scale=0.5]{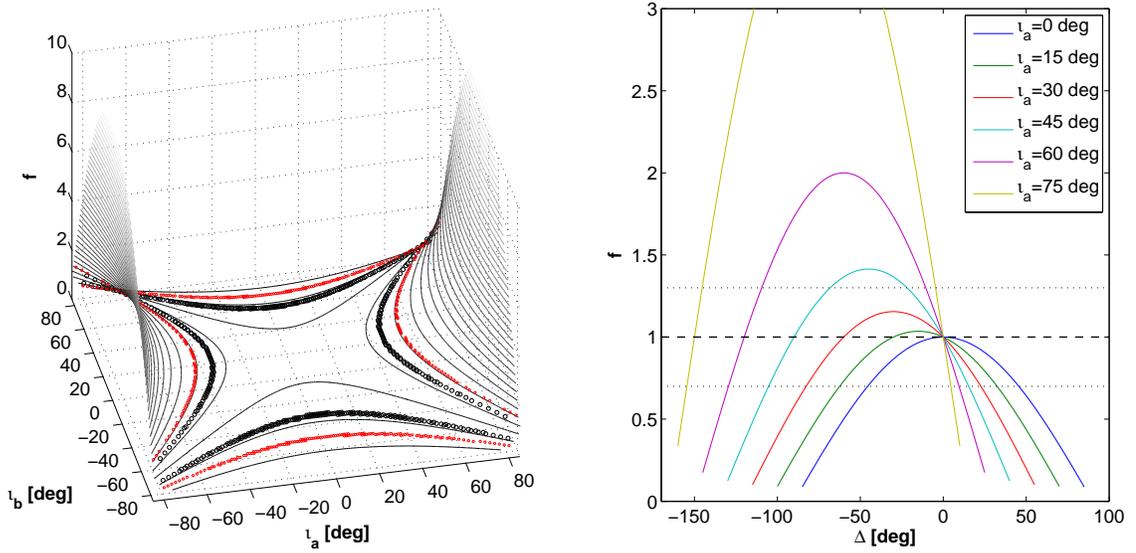}
 \caption{\label{projection_effect}
Left Panel: The projection function $f(\iota_a,\iota_b)=\frac{\cos \iota_b}{\cos\iota_a}$ 
for the two inclination angles $\iota_a$ and $\iota_b$ (Figure \ref{schematic_3d}).
The relatively flat saddle in the center shows that projection effects are small ($f\sim 1$) 
for $\iota_a$ and $\iota_b$ within about $\pm 40^{\circ}$.
Highlighted in black and red are the parameter spaces where the projection effects are 
within $\pm 30$\% ($f$ within $1\pm 0.3$) and  within $\pm 50$\% ($f$ within $1\pm 0.5$),
respectively.
For much larger inclination angles polarization is not easily detectable.
It is, thus, unlikely that severe
projection effects are present if regular polarization patterns are observed. 
Right Panel: The projection effect as a function of the difference $\Delta$ between the two
inclination directions, $\Delta=\iota_b-\iota_a$, for various fixed inclinations $\iota_a$.
The dashed line marks $f \equiv 1$. A $\pm 30$\% bound is shown with the dotted lines. 
}
\end{center}
\end{figure}

Strictly speaking, projection effects are likely to be present to some extent in all 
variables in Equation (\ref{B}). 
Having deprojected the intensity emission profile and adopted an average
field radius
(Section \ref{result_field_strength}), we limit the discussion to the remaining factor 
$\frac{\sin\psi}{\sin\alpha}$. In the following, 
lower indices {\it '3'} or {\it '2'} refer to 3-dimensional or 2-dimensional projected quantities. 
The deprojection of the side $b_2$ of the triangle (Figure \ref{schematic_3d})
can be written as $b_3=\sqrt{b_2^2+\pi_b^2}$ and $\tan \iota_b=\frac{\pi_b}{b_2}$, where
$\pi_b$ is the length of the projection along the line of sight ($z$-direction). Analogous
relations hold for the side $a_2$. Linking a sinus theorem in the 3D triangle, 
$\frac{\sin\psi_3}{\sin\alpha_3}=\frac{a_3}{b_3}$, to the previously used 2D expression
$\frac{\sin\psi_2}{\sin\alpha_2}=\frac{a_2}{b_2}$, then yields:
\begin{equation}
\frac{\sin\psi_3}{\sin\alpha_3}=\frac{\sin\psi_2}{\sin\alpha_2} \cdot
                                 \sqrt{\frac{1+\tan^2 \iota_a}{1+\tan^2 \iota_b}}
                           =\frac{\sin\psi_2}{\sin\alpha_2}\cdot \frac{\cos \iota_b}{\cos\iota_a}
\end{equation}
where $\iota_a,\iota_b \in [-\pi/2, \pi/2]$. It becomes obvious that for any inclination
angles, the projection effects cancel out as long as gravitational pull ($\iota_b$) and
field tension ($\iota_a$) are inclined by the same angles, i.e. $\iota_a \equiv \iota_b$.
Furthermore, if $\iota_a$ and $\iota_b$ are different, projection effects are negligible
as long as $\iota_a$ and $\iota_b$ are small. For $\iota_a$ and $\iota_b$ within about 
$\pm 40^{\circ}$, the projection effect is within about $\pm 20$\%. The left panel in 
Figure \ref{projection_effect} displays the relevant function $f(\iota_a,\iota_b)=
\frac{\cos \iota_b}{\cos\iota_a}$, which shows a large relatively flat saddle
around one
in the center.
For extreme values, i.e. inclination angles $|\iota_a|,|\iota_b| > 75^{\circ}$, the 
projection effect becomes severe ($f\rightarrow 0$ or $f\rightarrow \infty$).
Besides having two totally different inclination angles $\iota_a$ and $\iota_b$,
we also investigate the projection effect as a function of the difference between the 
two inclination directions, $\Delta=\iota_b-\iota_a$ (Figure \ref{projection_effect},
right panel). Since the gravitational pull and 
the field tension directions are dynamically coupled, their inclination angles are likely
to be not completely independent. The right panel in Figure \ref{projection_effect} shows
that projection effects are within about $\pm 30$\% for a range in the differences $\Delta$
up to about 100$^{\circ}$ or more, up to inclinations of about $\pm 45^{\circ}$.
For larger inclinations $f$ grows again. These situations, however, might not be of a 
concern, because at such large inclinations of the field lines, polarization emission
(perpendicular to the field lines) might not be detectable. 
The analysis might, therefore, not be applicable.

In summary, for identical inclination angles projection effects cancel out. The effect is 
small (up to $\sim 30$\%) or even negligible for one inclination up to $\sim 45^{\circ}$ 
and the other one varying within about $+50^{\circ}$ and $-50^{\circ}$ (range in $\Delta$
up to about $100^{\circ}$). Only for extreme inclinations larger than about $75^{\circ}$
projection effects can severely affect the method.
Such situations might, however, not be detectable. 
It, thus, has to be emphasized that it is unlikely that severe
projection effects are present if regular polarization patterns are observed. 
Consequently, the method introduced here leads to a total magnetic field strength.

\subsection{Shortcomings of the Method}

Without any further modeling, the method does not include components such as 
e.g., rotation or temperature variations. Effects due to rotation are not explicitly
considered in Equation (\ref{mhd_momentum}). Nevertheless, an additional term
due to rotation on the right hand side of Equation (\ref{mhd_momentum}) would 
primarily lead to a change in the resulting direction of motion, 
$\rho v \frac{\partial v}{\partial s_v}\mathbf{e}_{s_v}
+\rho v^2 \frac{\partial\mathbf{e}_{s_v}}{\partial s_v}$ in Equation (\ref{mhd_2}),
which was identified with the intensity gradient direction. Technically, this is 
equivalent to an additional inclination in the intensity gradient and can thus
be approximated as a projection effect. Similarly, the directions of the magnetic
field lines dragged along with the rotating gas would be subject to an additional
projection effect. As discussed in Section \ref{projection}, the resulting 
uncertainty in the field strength is relatively small for a large parameter space
of inclination angles. Alternatively, if the rotation velocity
and axis are
known, Equation 
(\ref{mhd_momentum}) could be written in a rotating frame
with an additional contribution from a centrifugal force.

Density and temperature combined make up for the detected dust emission. Significant
variations in temperature could mimic mass concentrations which might then 
lead to wrong gravity directions and, therefore, to wrong magnetic field strength values.
If concentrated locally such as e.g., in the case of radiation pressure from a star, 
this is unlikely to alter the global features in the field strength over an entire map.
Similarly, small shifts in the position of a dominating emission peak (like in 
Figure \ref{w51_e2}) introduce only minor variations in the field strength. Only if 
there are very irregular and significant changes in temperature over a map, the method
is likely to fail. Additional information (such as from spectra) might then help 
to isolate the temperature effect.

\subsection{Dynamical Importance of the $B$ Field: 
Flux-Freezing and the Magnetic Field - Intensity Gradient Relation}
\label{dynamic_b}

Following up on Section \ref{application}, we expect the strong and weak 
field case of flux-freezing to be present with different characteristics.
In the strong field case -- gas moving along the field lines -- the 
intensity gradient direction is presumably closely aligned with the 
magnetic field
orientation.
On the other hand, in the weak field case --
field lines dragged along and shaped by the forces acting on the gas --
intensity gradient and field
orientations
can be at any angle. In the 
extreme case they can be at a 90$^{\circ}$ angle, when e.g. a gas flow
is bending a field line (case (IIc) in Figure \ref{qualitative}). 
The angle $\delta(=\pi/2-\alpha)$, as previously 
introduced in Figure \ref{schematic_b}, measures the absolute difference in  
orientations
between the intensity gradient and the magnetic field. 
Since the cases of weak and strong\footnote{
We remark that the terms 'weak' and 'strong' case of flux-freezing are relative to 
the other competing forces (gravity, ram pressure, etc.). They are further
limited to a local region where these forces are evaluated on a map. 
Therefore, the $B$ field in a weak field region is not necessarily weaker than 
the field in a strong field zone.
} 
flux-freezing categorize the relative importance 
of the field, we propose here the angle $\delta$ as a diagnostic measure
for the dynamical role of the $B$ field. Small angles ($\sim 0$) will
point toward a dominating field, whereas larger and larger angles mean
that the field is being overwhelmed by other forces.

The map in Figure \ref{flux_freezing} displays $\delta$, 
revealing some areas with $\delta\approx 0$. These areas roughly
coincide with the expected hourglass at larger radii. Here, the gas
is possibly collapsing along the field lines. 
Additionally, it is likely that gravity has initially led to this field 
and intensity gradient configuration, with $\delta\approx 0$ being a mere
consequence of that. Both at smaller
radii toward the center and away from the hourglass toward the 
accretion plane, values for $\delta$ increase. This might be an 
indication that the field is dynamically less important here and that
gravity has largely taken control. 
It needs to be stressed that this does not necessarily mean that the field 
strength is smaller here, because $B$ further also depends on $R$, $\nabla \phi$
and $\nabla P$ (see also previous footnote). Moreover, the field can still 
provide pressure support which can slow down the gravitational collapse.
This picture -- with generally smaller $\delta$ values around the core and larger
values toward the north-west corner -- is consistent with the angle factor
$\frac{\sin\psi}{\sin\alpha}$ (Section \ref{local_field_importance} and top panels 
in Figure \ref{local_factors}),
which identifies the field force to be less significant in the core.
Figure \ref{flux_freezing} is necessarily complementary to the middle panels
in Figure \ref{angles}, because $\delta=\pi/2-\alpha$. 

We have to note that the angle $\delta$ (like $\alpha$) is subject to a projection 
correction. In the notation of Section \ref{projection}: $\delta_3=\pi/2-\alpha_3$
with $\sin\alpha_3=\frac{\sin\psi_3}{\sin\psi_2}\cdot \sin\alpha_2\cdot f^{-1}(\iota_a,\iota_b)$.
In the particular case of a close to spherical collapse, $\frac{\sin\psi_3}{\sin\psi_2}\approx 1$,
because both the 3-dimensional and the projected intensity show a clear azimuthal
symmetry (Figure \ref{w51_e2}). Assuming $f(\iota_a,\iota_b)$ to be approximately
constant over a map (Section \ref{projection}), then gives $\sin\alpha_3\approx const\cdot\sin\alpha_2$.
Thus, despite the projection effect, the dynamical importance of the 
field can still be understood from the relative differences in $\delta$.
The significance of the angle $\delta$ as a diagnostic measure is further 
analyzed on a larger data set together with the correlation from 
Section \ref{motivation} in Koch et al., 2012 (in preparation).

\subsection{Magnetic Field and Gravity Alignment: $\sin\omega$ -- map}
\label{b_gravity}

\begin{figure}
\begin{center}
\includegraphics[scale=0.65]{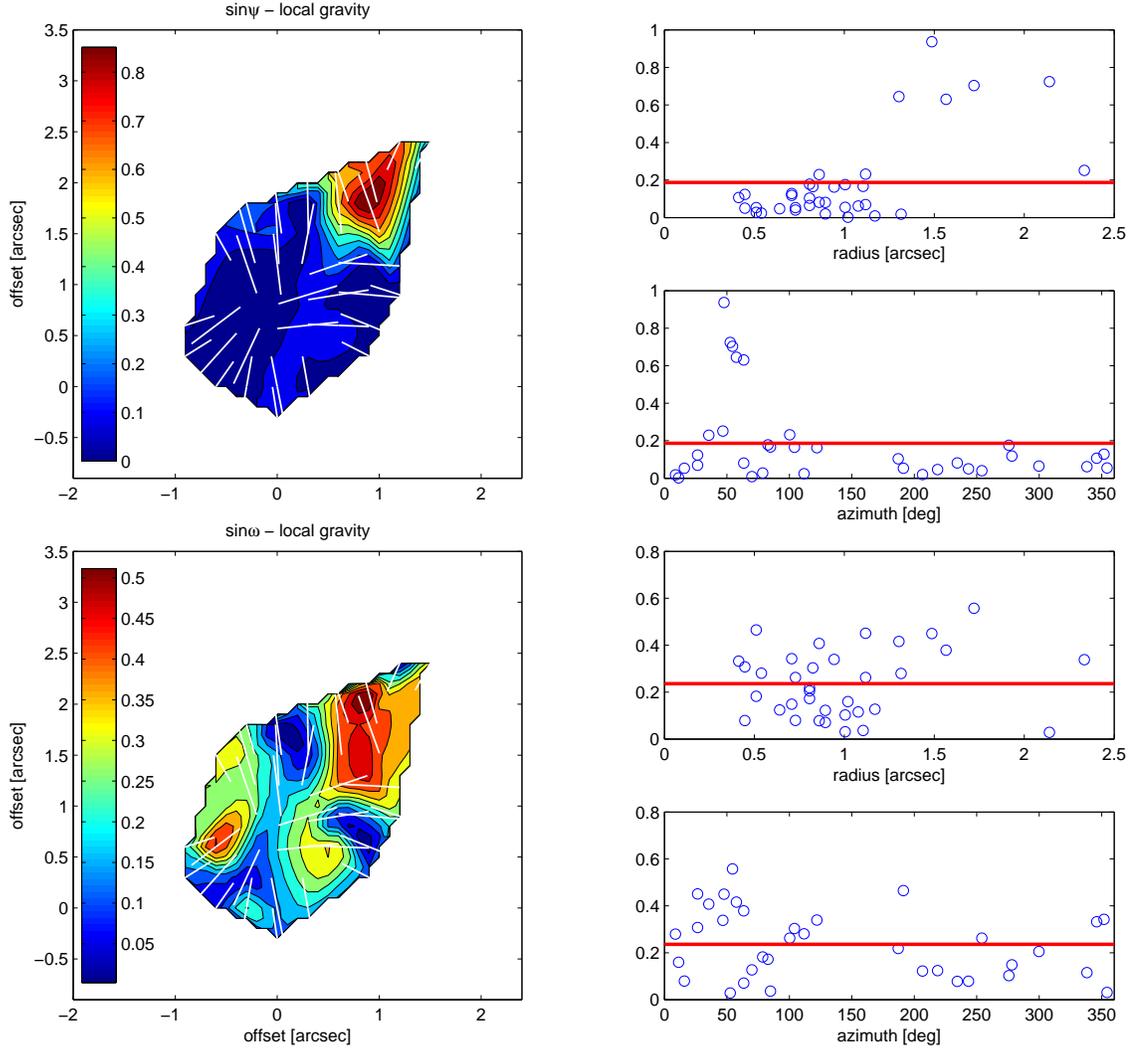}
 \caption{\label{angles_local_gravity} \footnotesize
Angles corrected for local gravity directions.
Magnetic field segments are shown in white.
Top panels: $\sin\psi$ measuring the deviation between 
the local gravity and the intensity gradient 
orientations.
Bottom panels: $\sin\omega$ measuring the deviation between 
the local gravity and the magnetic field 
orientations.
}
\end{center}
\end{figure}

\begin{figure}
\begin{center}
\includegraphics[scale=0.7]{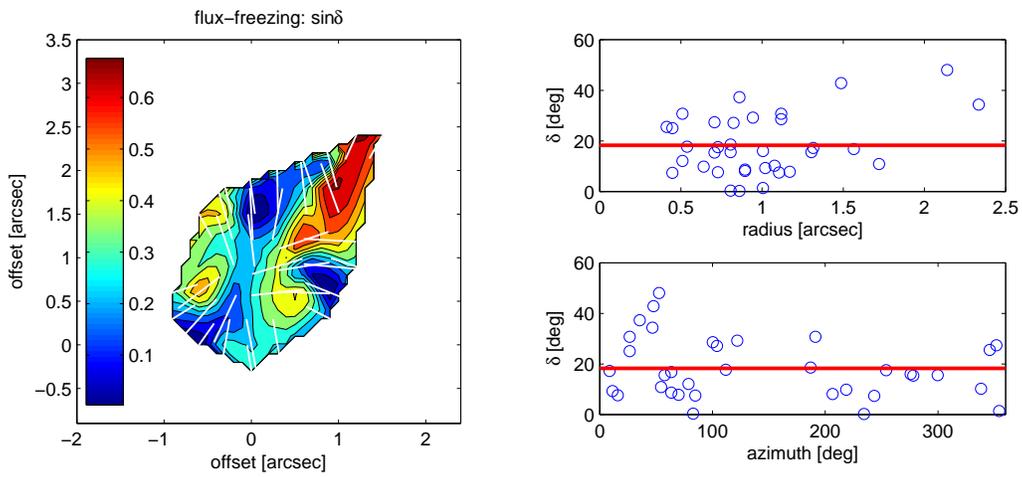}
 \caption{\label{flux_freezing}
The angle $\delta$ (absolute difference in between the intensity gradient
and the magnetic field 
orientations
(Figure \ref{schematic_b})) 
as a diagnostic tool to distinguish between weak and 
strong case of flux-freezing. 
Magnetic field segments are shown in white.
The values for $\sin\delta$ are displayed in the left panel in order to facilitate a 
comparison with the $\sin\alpha$ (Figure \ref{angles}, middle panels) and 
the $\sin\omega$ maps (Figure \ref{angles}, bottom panel; 
Figure \ref{angles_local_gravity}, bottom panel).}
\end{center}
\end{figure}

The method developed in Section \ref{method} is based on identifying the 
three 
orientations
of gravity, the magnetic field and the emission intensity 
gradient. In particular, two angles are relevant in order to isolate and 
calculate the magnetic field strength $B$: the angle $\psi$ 
between the gravity direction and the intensity gradient, and the angle 
$\alpha$ between the polarization 
orientation
and the intensity gradient. 
Given the three 
orientations,
an additional angle, $\omega$, can be read off
from a map. $\omega$ measures the difference in 
orientations
between gravity and
the magnetic field (Figure \ref{schematic_b}). Thus, it is a measure for how
effectively gravity is able to shape the field lines and aligning them with 
the direction of the gravitational pull. Small angles (small values of 
$\sin\omega$) indicate close alignment. The bottom panel in Figure \ref{angles}
shows $\sin\omega$ for the case of W51 e2, assuming the center of gravity
to be at the emission peak. The bottom panels in Figure \ref{angles_local_gravity} 
display again $\sin\omega$ when a local gravity direction 
(Section \ref{local_gravity}) is adopted. The two figures are qualitatively 
similar. Applying a local gravity direction seems to sharpen the contrasts.
The closest alignment 
($\sin\omega \approx 0.1-0.2$) is found at the outer radii close to the 
accretion plane in the west and in the north-west direction around the 
axis of the hourglass-like structure. An obvious poor alignment is 
in the northern area. Averaged over the entire map and excluding the 
northern patch we derive $<\sin\omega>\approx 0.21$ ($<\omega>\approx 13^{\circ}$)
and $<\sin\omega>\approx0.14$ ($<\omega>\approx 8^{\circ}$), respectively.
Interestingly, after the local gravity correction, there is a very close
resemblance between the $\sin\omega$-map and the $\sin\delta$-map
(Figure \ref{flux_freezing}). Since $\omega$ measures the magnetic field - 
gravity alignment and $\delta$ measures the magnetic field - intensity 
gradient alignment, the very similar maps, of course, indicate that the 
intensity structure is mostly resulting from gravity. This is also shown
with the small values in the $\sin\psi$-map (Figure \ref{angles_local_gravity},
top panels). This is, of course, a consequence of the geometry in Figure 
\ref{schematic_b} because all the 3 angles are dependent: $\omega=\pi/2-\alpha-\psi$
with $\alpha=\pi/2-\delta$. Nevertheless, the clear systematic differences
in structures between the $\sin\omega$ and the $\sin\psi$-map (Figure 
\ref{angles_local_gravity}) lead to the conclusion that the $B$ field -- at 
least in some regions -- has kept its own dynamics and is not yet aligned
and overwhelmed by gravity. This result is then also consistent with the 
finding of zones of weak and strong flux-freezing (Section \ref{dynamic_b})   
which characterizes the dynamical role of the magnetic field.  
The systematic influence of the field is further manifest in the systematic
deviations from an azimuthally symmetric emission structure, seen in the 
bottom panels in Figure \ref{local_factors}.
Projection effects (Section \ref{projection}) are again likely to change
only the absolute numbers, leaving relevant features and comparisons between
maps still intact.

\section{Summary and Conclusion}  \label{summary}

Dust polarization observations in molecular clouds often show position angles of magnetic 
field segments which are close to perpendicular to the dust continuum intensity contours.
This correlation does not seem to be arbitrary, but can be interpreted in the context of 
ideal magneto-hydrodynamics (MHD). Based on this, we put forward a new method to derive a 
local magnetic field strength. The key points are summarized in the following.

\begin{enumerate}

\item 
{\it New method for magnetic field strength mapping:} 
A new method is proposed which leads to a position-dependent estimate of the magnetic 
field strength. The approach is based on measuring the angle between the polarization 
and the Stokes $I$ intensity gradient
orientations.
Assuming that the intensity gradient 
is a measure for the resulting direction of motion in the MHD force equation, this angle
quantifies how tightly the direction of motion is coupled to the field
orientation.
In 
combination with the angle in between the 
orientations
of gravity (and/or the pressure 
gradient) and the intensity gradient, the two angles can be linked to the field strength
by identifying a force triangle.

\item
{\it Application regime:}
The method relies on separating noticeable effects from various components in the 
MHD force equation, in particular from gravitational pull and field tension.
Consequently, the method is more directly applicable to the case of weak field flux
freezing, where the field morphology is clearly shaped by these forces.
Collapsing (spherical) cores in star formation sites are of most immediate interest
because the directions of gravity and magnetic field are relatively easily determined.
Nevertheless, the technique is generally valid and applicable to any measurement
(also strong field flux freezing), where the directions of magnetic field, pressure
gradients and/or gravitational pull can be localized. With the concepts of local
curvature and local gravity direction, the method can be further generalized.
Moreover, the method is not limited to dust polarization measurements, but can 
potentially also be applied to, e.g. molecular line linear polarization maps from the 
GK effect or even to Faraday rotation measure maps for galaxies. 

Here, the method is applied to SMA
dust
polarization data of the collapsing core W51 e2. 
Averaging over the entire core, a field strength of $\sim 7.7$~mG is found. 
Variations in the field strength are detected with an azimuthally averaged
radial profile $B(r)\sim r^{-1/2}$.
Maximum values toward the center are about $19$~mG.
The current data lack resolution 
in order to probe the innermost part of the core, where the largest
field strength is expected.

\item
{\it Comparison with Chandrasekhar-Fermi (CF) method:}
The method introduced here uses the large-scale field curvature. The local turbulent 
field dispersion is acting as a contaminant. This is in contrast to the CF method which
precisely uses the field dispersion -- after separating a mean field
orientation -- in 
combination with a turbulent velocity information in order to derive a field strength.
Whereas the CF method yields one statistically averaged field strength for an
ensemble of polarization segments, the method here provides a field strength at each 
location of a detected polarization segment.

\item
{\it Limitations and shortcomings of the method:}
Technically, the method fails where the force triangle can not be closed. This leads 
to the unphysical value $B\rightarrow \infty$. A triangle degenerating to a single
line corresponds to the situation where the field is very weak and, therefore, the 
intensity gradient does not deviate from the gravity direction. The method then still
consistently yields the limit $B\rightarrow 0$.

The method is not entirely free of projection effects. In the most general case, the 
force triangle has to be deprojected with two different inclination angles. Nevertheless,
the projection correction enters as the ratio of the two inclination angles. Thus, in the 
most favorable case with identical inclination angles for gravity and field
orientations,
the projection effect cancels out and the method leads to a total magnetic field
strength. Even if the inclination angles differ from each
other, the possible error due to unknown inclinations is relatively modest, within 
$\pm 30$\% to $\pm 50$\%, for a large parameter space. 
In any case, if the inclination angles do not vary significantly within the same source, 
the method always provides a way to detect relative differences in the field strength.
The accuracy and reliability of 
the method are limited by how well projection effects and forces defining the field
morphology can be identified and quantified. Additional effects like, e.g. 
rotation and local variations in temperature are likely of minor importance.

\item
{\it Weak and strong flux freezing:}
The dynamical role of the magnetic field -- in the sense of weak or strong flux 
freezing -- can be assessed through the angle between the magnetic field and the 
intensity gradient
orientations.
Angles close to zero point toward a dominating role
of the field (or an overall controlling gravitational pull). Larger angles indicate
that the field lines are more significantly shaped by other forces. Although
this criterion alone does not constrain the field strength, 
it provides a measure to characterize the role of the magnetic 
field as a function of position in a map.

\item
{\it Dynamical significance of magnetic field:}
The method additionally provides a model-independent criterion to quantify the 
magnetic field force compared to the sum of any other forces involved. As a result 
of the force triangle, the local 
significance
of the magnetic field can be 
evaluated simply with the ratio of the angle between intensity gradient and gravity
direction, and the angle between intensity gradient and polarization
orientation. These angles reflect the imprint of the various forces onto the dust and 
polarization morphologies.
This measure can then constrain the local impact of the magnetic field and quantify its
support against gravitational collapse.

\end{enumerate}

The authors thank the referee for valuable comments which led to further
significant insight in this work.
P.T.P.H. is supported by NSC grant NSC97-2112-M-001-007-MY3.




\begin{deluxetable}{ccccccc} \rotate
 \tabletypesize{\scriptsize}
\tablewidth{0pt}
\tablecaption{Qualitative Analysis\label{table_analysis}}
\tablehead{ \colhead{} &
 \colhead{$\psi$} &  \colhead{$\alpha$} &  \colhead{$R$} & \colhead{$\Sigma_B$} & \colhead{$B$} & \colhead{Comments} \\
}
\startdata
General\\
\cline{1-1}\\
&$\downarrow \uparrow$  & $\neq 0$   & $R$ & $\downarrow \uparrow$ & $\downarrow \uparrow$, $\sim\sqrt{\sin\psi\cdot R}\cdot r^{-\kappa}$  
         &  $\Sigma_B$ and $B\uparrow$ with larger angle $\psi$  \\
&$\neq 0$         & $\downarrow \uparrow$ & $R$ & $\uparrow \downarrow$ & $\uparrow \downarrow$, $\sim\sqrt{\frac{R}{\sin\alpha}}\cdot r^{-\kappa}$ 
        &  $\Sigma_B$ and $B\uparrow$ with smaller angle $\alpha$ \\
  &&&&&&\\

Specific Cases\\
\cline{1-1}\\
(I) (a)  & small, $\approx$ const  & $\sim \frac{\pi}{2}$  & $\approx$ const  & $\approx$ const  & $\approx$ const$\cdot r^{-\kappa}$
         & field line close to pole, $\Sigma_B\approx$ const, $B\uparrow$ with smaller $r$ \\
    (b)  & small, $\approx$ const  & $\sim \frac{\pi}{2}$  & $\approx$ const  & $\approx$ const  & $\approx$ const$\cdot r^{-\kappa}$  &\\
(II) (a) & small, $\approx$ const  & $\sim \frac{\pi}{2}$  & $\approx$ const  & $\approx$ const  & $\approx$ const$\cdot r^{-\kappa}$ 
         & field line around accretion direction, $\Sigma_B\approx$ const, $B\uparrow$ with smaller $r$ \\
    (b)  & small, $\approx$ const  & $\downarrow$          & $\downarrow$     & $\sim\frac{1}{\sin\alpha}$    & $\sim\sqrt\frac{R}{\sin\alpha}\cdot r^{-\kappa}$  
         & $\Sigma_B\uparrow$, $B\uparrow$ with smaller $r$ steeper or shallower than $r^{-\kappa}$  (depending on $R$ and $\alpha$)\\ 
    (c)  & small, $\approx$ const  & $\downarrow$, $\sim 0$& $\downarrow$, $R_{min}$ & $\sim\frac{1}{\sin\alpha}$ &   $\sim\sqrt\frac{R}{\sin\alpha}\cdot r^{-\kappa}$  
         & $\Sigma_B\uparrow$, maximum field strength $B$\\ 
(III)      & $\uparrow\downarrow$    &  $\approx$ const      &  $\approx$ const     & $\uparrow\downarrow$    & $\uparrow\downarrow$, $\sim\sqrt{\sin\psi}\cdot r^{-\kappa}$ 
         & outer area, no clear signatures yet, $B$ generally small due to $r^{-\kappa}$ \\
(IV)     & large                   &  $\approx$ const      & $\approx$ const     & large    & $\sim\sqrt{\sin\psi}\cdot r^{-\kappa}$ 
          & disconnected area, no clear gravity center localized, $B$ small due to $r^{-\kappa}$\\
  &&&&&&\\

Limits\\
\cline{1-1}\\

&$\rightarrow 0$      &  $\neq 0$               & (i) $R^1$, (ii) $\rightarrow\infty^{2}$ & $\rightarrow 0$  
        & (i) $\rightarrow 0^1$, (ii) $\rightarrow\infty^{2}$  
        & triangle degenerates to single line\\
&$\neq 0$            &  $\rightarrow 0$         & (i) $R$, (ii) $\rightarrow\infty$ & $\rightarrow\infty$  
        & (i) $\rightarrow\infty$, (ii) $\rightarrow\infty$ 
        & method fails, triangle not closed \\
&$\rightarrow 0$      &  $\rightarrow 0$        & (i) $R$, (ii) $\rightarrow\infty$ & $\rightarrow$1  & (i) $\sim\sqrt{R}$, (ii) $\rightarrow\infty$ 
        & method fails, triangle not closed and degenerated to single line   \\

\enddata
\tablecomments{Qualitative analysis of the force ratio $\Sigma_B=\frac{\sin\psi}{\sin\alpha}$ and the 
magnetic field strength $B=\sqrt{\frac{\sin\psi}{\sin\alpha}\rho\nabla\phi\cdot 4\pi R}$ 
(Section \ref{analysis}). 
The pressure gradient is neglected here.
The profile of the density and the resulting gradient of the gravitational potential is represented
with $r^{-\kappa}$ with $\kappa > 0$ (Section \ref{result_field_strength}).
The specific cases labeled (I) to (IV) refer to Figure \ref{qualitative}.
The cases (Ib), (IIb) and (IIc) are not yet observed due to the lack of higher resolution data. 
Down-arrows ($\downarrow$) and up-arrows ($\uparrow$)
indicate decreasing and increasing values, respectively. 
'$\approx$const' indicates that small local variations can occur which lead to some scatter in $\Sigma_B$ and $B$
(Figure \ref{b_map_local_profile}).
Failure of the method -- triangle 
can not be closed -- refers to Figure \ref{schematic_b} .
}
\tablenotetext{1}{The case $\psi\rightarrow 0$ states generally that the intensity gradient does not deviate from 
the gravity direction for any orientation of the field lines. Therefore, the field strength must be insignificant,
i.e. $B\rightarrow 0$, which is also consistently reproduced by the method. 
}
\tablenotetext{2}{The case $\alpha\neq 0\sim \pi/2$ with $R\rightarrow\infty$ -- i.e., straight B field lines aligned 
with the intensity gradient -- represents the strong field flux freezing situation (see Section \ref{application})
and consistently yields a dominating field strength ($B\rightarrow\infty$, $\Sigma_B\rightarrow\infty$).}
\end{deluxetable}

\end{document}